\documentclass[pdflatex,sn-nature,iicol]{sn-jnl}

\usepackage{graphicx}%
\usepackage{multirow}%
\usepackage{amsmath,amssymb,amsfonts}%
\usepackage{amsthm}%
\usepackage{mathrsfs}%
\usepackage[title]{appendix}%
\usepackage{xcolor}%
\usepackage{textcomp}%
\usepackage{manyfoot}%
\usepackage{booktabs}%
\usepackage{algorithm}%
\usepackage{algorithmicx}%
\usepackage{algpseudocode}%
\usepackage{listings}%

\usepackage{type1cm} %
\usepackage{xr} %
\usepackage{filecontents}

\unnumbered%

\graphicspath{{./figures/}}

\begin{document}

% work around broken cross-references between main article and SI document
\makeatletter\@input{supplementary_information_xr_references.tex}\makeatother

\title[Article Title]{In silico high-resolution whole lung model to predict the locally delivered dose of inhaled drugs}

\author*[1]{\fnm{Maximilian J.} \sur{Grill}}\email{grill@ebenbuild.com}

\author[1]{\fnm{Jonas} \sur{Biehler}}

\author[1]{\fnm{Karl-Robert} \sur{Wichmann}}

\author[1]{\fnm{David} \sur{Rudlstorfer}}

\author[1]{\fnm{Maximilian} \sur{Rixner}}

\author[1]{\fnm{Marie} \sur{Brei}}

\author[1]{\fnm{Jakob} \sur{Richter}}

\author[1]{\fnm{Joshua} \sur{B{\"u}gel}}

\author[1]{\fnm{Nina} \sur{Pischke}}

\author[]{\fnm{Wolfgang A.} \sur{Wall\textsuperscript{1,2}}} %

\author[1]{\fnm{Kei W.} \sur{M{\"u}ller}}

\affil[1]{\orgname{Ebenbuild GmbH}, \orgaddress{\city{Munich}, \country{Germany}}}

\affil[2]{\orgdiv{Institute for Computational Mechanics}, \orgname{Technical University of Munich}, \orgaddress{\country{Germany}}}

\abstract{
The big crux with drug delivery to human lungs is that the delivered dose at the local site of action is unpredictable and very difficult to measure, even a posteriori.
It is highly subject-specific as it depends on lung morphology, disease, breathing, and aerosol characteristics.
Given these challenges, computational approaches have shown potential, but have so far failed due to fundamental methodical limitations.
We present and validate a novel in silico model that enables the subject-specific prediction of local aerosol deposition throughout the entire lung.
Its unprecedented spatiotemporal resolution allows to track each aerosol particle anytime during the breathing cycle, anywhere in the complete system of conducting airways and the alveolar region.
Predictions are shown to be in excellent agreement with in vivo SPECT/CT data for a healthy human cohort.
We further showcase the model's capabilities to represent strong heterogeneities in diseased lungs by studying an IPF patient.
Finally, high computational efficiency and automated model generation and calibration ensure readiness to be applied at scale.
We envision our method not only to improve inhalation therapies by informing and accelerating all stages of (pre-)clinical drug and device development, but also as a more-than-equivalent alternative to nuclear imaging of the lungs.
}


\maketitle

With an average development cost of \$115 million and a bench-to-bedside success rate of only 16\%, developing novel inhaled therapies for the more than 500 million patients \cite{sorianoPrevalenceAttributableHealth2020} suffering from chronic respiratory diseases (CRDs) is very expensive, difficult and risky.
One common difficulty that arises in the development of locally acting medications for CRDs like Asthma, COPD, or IPF is the quantification and optimization of the so-called local effective dose at the side of action, i.e., the amount of active ingredient that reaches the targeted areas within the lungs.
Hence, aerosol transport and deposition and in turn the local effective dose depend not only on inhaler design and inhalation pattern, aerosol type and particle size distribution that can be influenced by the drug or device manufacturer, but also on patient dependent factors such as constitution, or anatomical alterations like dysanapsis~\cite{smithHumanAirwayBranch2018}, and the disease state.
As a result, obtaining proof of performance for a delivery device or the bioequivalence of generics is often very difficult.
In this paper, we propose a novel in silico approach that can overcome the aforementioned challenges and provide the first cost-effective means of optimizing pulmonary drug delivery with respect to delivery device and aerosol parameters.

To date, researchers need to rely on a combination of in vitro, ex vivo, in vivo and in silico studies to obtain a comprehensive understanding of the beneficial as well as detrimental effects of an inhaled drug.
While in vitro testing is the foundation, studies are limited to either standardized test systems such as cascade impactors~\cite{olssonVitroMethodsStudy2021} or simplified, physical lung models~\cite{delvadiaVitroTestsAerosol2012} and even then these experiments are notoriously difficult to perform.
Ex vivo and in vivo animal tests are the next step.
In addition to ever-increasing ethical concerns, developing appropriate animal models is very challenging, and no animal perfectly mimics human disease, physiology, anatomy and, perhaps most importantly, any clinically relevant mode of inhaled drug delivery
\cite{secherCorrelationClinicalRelevance2020}.
As a result, any information about the local effective dose that is obtained in animal studies is difficult to transfer or extrapolate to human subjects.
In vivo human studies also often provide little or no information on the local effective dose.
One exception is the use of nuclear imaging techniques which makes in vivo imaging of aerosol inhalation and deposition in humans possible through radioscintigraphy or single photon emission computed tomography (SPECT)~\cite{conwayControlledParametricIndividualized2012,majoralControlledParametricIndividualized2014, flemingControlledParametricIndividualized2015}.
Even though spatial and temporal resolution of these imaging techniques is limited, SPECT images can be used to quantify aerosol deposition pattern including its regional distribution, e.g., per lobe or by dividing the lungs into a central and peripheral region to obtain a proxy for the amount of drug deposition in the respiratory airways. 
However, its exposure of the patient to radiation and the associated high cost limits its widespread use, even within the scope of clinical studies.  

In silico modeling in principle promises to supplement experimental information and remedy some of the fundamental shortcomings of the aforementioned experimental methods.
However, the enormous complexity of the lungs with typically tens of thousands of conducting airways~\cite{haefeli-bleuerMorphometryHumanPulmonary1988} and half a billion alveoli~\cite{ochsNumberAlveoliHuman2004b} has limited in silico investigations to either truncated or otherwise simplified model systems and so far prevented the patient-specific, organ-level in silico analysis of pulmonary drug delivery.
And hence the challenge of accurately computing the local effective dose at the site of action is still unmet.

Existing empirical and heuristic lung models are unsuitable for simulating patient-specific aerosol transport and deposition due to their lack of personalization and inability to account for airway or alveolar tissue pathology. 
This applies to mathematical filter-in-series models such as the ICRP model~\cite{baileyNewICRPModel1994} as well as the seminal Trumpet model~\cite{taulbeeTheoryAerosolDeposition1975, taulbeeAerosolTransportHuman1978, lazaridisIntegratedExposureDose2001, robinsonDepositionCigaretteSmoke2001, mitsakouEulerianModellingLung2005, choiMathematicalAnalysisParticle2007a} which approximates the human respiratory system through a one-dimensional variable cross-section channel and therefore neglects the entire internal geometry and structure of the lung.

Morphometric models such as single-~\cite{HumanRespiratoryTract1994a, weibelMorphometryHumanLung1963,yehModelsHumanLung1980} or the growing variety of multiple-path models~\cite{anjilvelMultiplePathModelParticle1995a, asgharianParticleDepositionMultiplePath2001, millerImprovementsAdditionsMultiple2016a, olssonMimetikosPreludiumNew2018a}, while based on the geometry of the airway tree, calculate the distribution of airflow within the lung not based on physical principles but on heuristic assumptions: flow splits proportionally to the distal lung volume supplied by each daughter airway.
This assumption is almost certainly incorrect once the respiratory system is diseased and shows pathological abnormalities.
All aforementioned approaches are only useful if applied to large cohorts of relatively healthy lungs to gain a general understanding of pulmonary drug delivery.

By contrast, physics-based 3D computational fluid particle dynamics (CFPD) models~\cite{faizalComputationalFluidDynamics2020,kleinstreuerAirflowParticleTransport2010,longestSilicoModelsAerosol2012a,debackerValidationComputationalFluid2010a} that solely consist of image-extracted airway tree geometries can provide patient-specific results.
Still, they have considerable drawbacks limiting their deposition predictions to coarse regional analysis such as the five lobes.
Regional deposition in central or peripheral regions can only be estimated making simplifying assumptions.
The computational domain on which the governing equations for air and particle dynamics are solved is typically limited to the first six to seven airway generations due to the limited image resolution of, e.g., a CT scan.
Such models effectively comprise a system of inelastic tubes with multiple outlets at the lobar or sublobar bronchi.
Smaller distal airways, the alveolar region, the pleura, diaphragm and the rib cage are omitted~\cite{debackerValidationComputationalFluid2010a,vanholsbekeUseFunctionalRespiratory2018,usmaniPredictingLungDeposition2021}. 
As a result, these approaches cannot model exhalation and the predicted deposition statistics are distorted as all inhaled particles are assumed to deposit upon passing through the outlets.
Central versus peripheral deposition can only be estimated via the particle fraction deposited in the first few generations and the total number of inhaled particles, which neglects higher generation airways that extend into the central region.
Further regional information beyond the lobar level or reliable deposition statistics for higher generations cannot be computed.

In the past, the concept of CFD and CFPD analysis was extended to airway generations below the resolution of in vivo CT-scans by augmenting the airway segmentation using morphological, space-filling trees~\cite{choiNumericalStudyHighFrequency2010, kronbichlerNextgenerationDiscontinuousGalerkin2021, zhaoPredictionCarrierShape2022, williamsEffectPatientInhalation2022,williamsValidatedRespiratoryDrug2023,oakesAirflowSimulationsInfant2018}.
While including airways beyond the seventh generation, such models still suffer from the drawbacks of either open outlets at the bronchioli or otherwise heuristically determined outlet boundary conditions using, e.g., lumped parameter models~\cite{rothComputationalModelingRespiratory2019}.
As a result these models cannot account for the crucial effects of diseased tissue on airflow distribution, which in turn governs aerosol transport and deposition.
In addition, the required computational resources can only be mustered by the largest existing high-performance computing systems.

So-called truncated whole lung models overcome this exponential complexity by resolving 
only a few arbitrarily selected branches of the airway tree up to generation 16-18, while truncating all the other airways earlier~\cite{waltersEfficientPhysiologicallyRealistic2011, zhaoPredictionAirwayDeformation2021}.
Although this approach allows, to a limited extent, for disease-specific mechanical properties of airways and an alveolar tissue model, the limited number of typical pathways from the upper airways to the alveoli precludes the ability to study the spatial distribution of aerosols within the lung. 

To the authors' knowledge, the approach described in~\cite{oakesAerosolTransportThroughout2017, poorbahramiWholeLungSilico2021} is the first to simulate exhalation by combining a 3D patient-specific CFPD model for the larger airways with 1D-trumpet models attached to its distal outlets.
Aerosol particles convected out of the 3D domain are accounted for as concentrations in the 1D trumpet models and seeded back into the 3D geometry during exhalation. 
While this approach considers the entire lung, the spatial resolution of aerosol deposition is still limited by the 3D model, and particle trajectories cannot be tracked beyond the 3D geometry.
The previously described shortcomings of the trumpet model make it difficult to include pathologies in smaller conducting airways and alveolar tissue.
A similar approach, subject to the same limitations, but using a different 1D formulation, was recently described in~\cite{kupratEfficientBidirectionalCoupling2021}.

In summary, all currently existing approaches that rely on 3D CFPD models have severe drawbacks that preclude them from accurately computing the transport and deposition of aerosol particles in the entire human lung, even more so if the lung exhibits pathologies.

In this paper, we present a novel in silico approach that provides unprecedented spatiotemporal resolution, allowing for each aerosol particle to be tracked at any time during the respiratory cycle and, for the first time, anywhere in the entire system of conducting airways and the alveolar region.
As a consequence, our model is the first effective method for calculating the local effective dose at the site(s) of action for pulmonary drug delivery.
We also present the results of a thorough validation study, comparing the model predictions with in vivo SPECT/CT measurements of drug deposition in a human cohort.
Finally, our simulation approach is embedded into a comprehensive computational framework that includes automated image processing, model generation, and calibration.
In contrast to the methodological proof-of-concept research mentioned above, which requires a lot of manual work and fine-tuning during model creation and simulation setup,   
we can create simulation-ready models from images in minutes and deliver final results and analysis in hours.
As a result, our approach is ready to be applied in in silico trials at scale.
This improvement represents an advancement of several orders of magnitude in both time-to-result and resolution.

\section{Results}\label{sec:results}

\begin{figure*}[htpb]%
  \centering
  \includegraphics[width=0.9\textwidth]{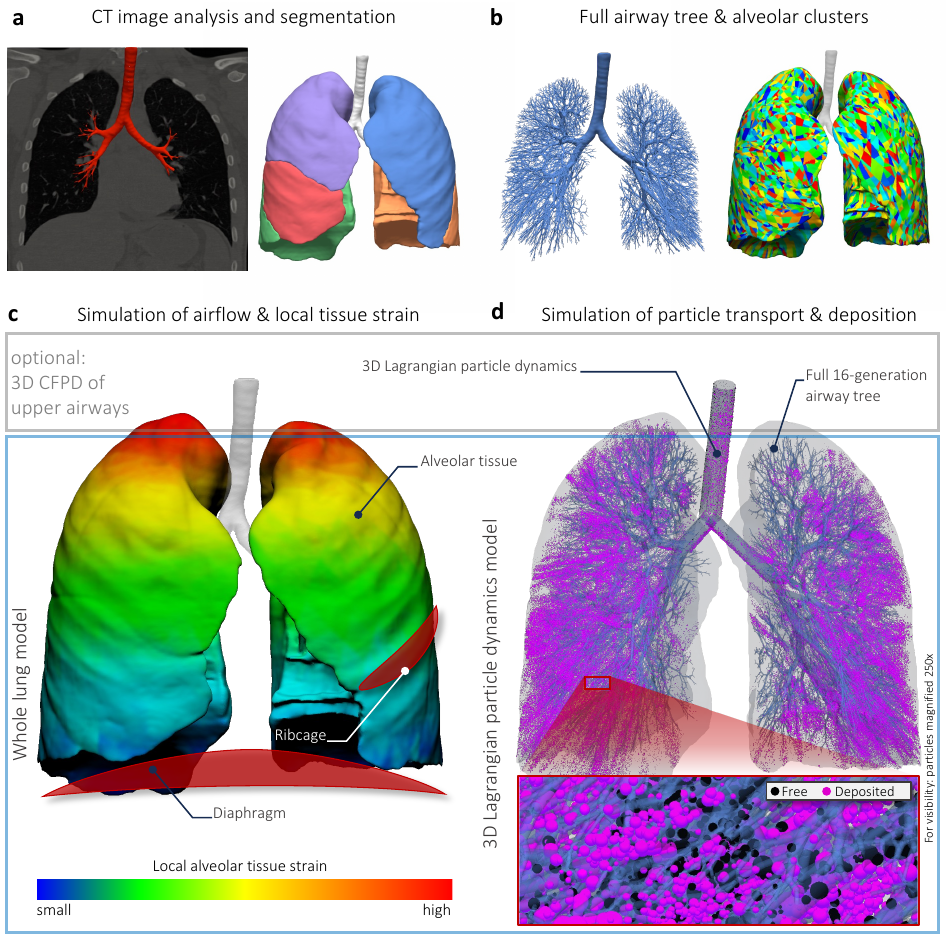}
  \caption{\textbf{Illustration of the model design and its subject-specific generation process} (here: subject H04).
      \textbf{a} Extraction of detailed volumetric information about lung, lobes, airways, and local gas volumes from the CT image.
      \textbf{b} Creation of the full 16 generation airway tree and alveolar clusters based on segmented volumetric information and a physiology-based space-filling growth algorithm.
      \textbf{c} Illustration of the reduced-dimensional whole lung model for the simulation of airflow and local tissue strain~\cite{ismailCoupledReducedDimensional2013,rothCouplingEITComputational2017,rothComprehensiveComputationalHuman2017}.
      \textbf{d} Particle transport and deposition model describing particle dynamics in a Lagrangian formulation in 3D throughout the entire whole lung model.
      Note that a 3D computational fluid and particle dynamics (CFPD) model of the upper airways can be coupled optionally.
      }
  \label{fig:model}
\end{figure*}

\bmhead{In silico model for respiratory drug delivery}
The components of our newly developed, comprehensive whole lung model are outlined in Fig.~\ref{fig:model}.
Its key idea is the first-of-its-kind combination of a 3D Lagrangian particle transport and deposition model with a previously published reduced-dimensional model for airflow and local tissue strain~\cite{ismailCoupledReducedDimensional2013, rothCouplingEITComputational2017,rothComprehensiveComputationalHuman2017}.
This way, we account for the full tree of conducting airways, the entire alveolar tissue (healthy and diseased), and additionally the effects of pleura, diaphragm and rib cage.
We can track individual particles at any time throughout inhalation and exhalation and at any point throughout the entire respiratory system which leads to an unprecedented level of physiological correctness and data resolution at the same time.
It is important to note that a 3D model of the upper airways including the mouth and throat can optionally be coupled to the reduced-dimensional whole lung model.
Due to the focus on the novel whole lung model and its validation, however, we will not consider this in the remainder of this work.
Instead we applied an analytical mouth-throat filter model~\cite{stahlhofenIntercomparisonExperimentalRegional1989}, which filters the aerosol particles based on deposition probabilities. 
As it turns out, this effective model already yields excellent validation results for the considered measures of aerosol deposition patterns within the lungs.
More details about the model and the numerical methods are provided in SI Sec.~\ref{sec:SI_model_methods}.

The generation of the subject-specific model starting from a single volumetric CT image is a streamlined and mostly automated process illustrated in Fig.~\ref{fig:model}.
The first step is to segment the geometry of the lungs, lobes and airways.
Once image resolution prohibits further extraction of the airway geometry, a morphology-based tree growth algorithm creates the remaining generations of conducting airways in a space-filling manner.
See SI Sec.~\ref{sec:SI_model_methods}\ref{sec:SI_model_gen} for details about the model generation process.
The subsequent flow simulation of a given arbitrary breathing pattern yields the distribution of airflow and local tissue strains throughout the entire respiratory system.
Based on this known flow field, the given aerosol specification and an arbitrary space-time-dependent aerosol mass rate, the particle simulation finally computes the full trajectory of each inhaled particle as it is transported through the airway tree and potentially deposits at an airway wall, in the respiratory zone, or is exhaled.
Possible post-processing steps include the analysis and the visualization of the resulting particle deposition pattern as will be demonstrated later.
To enable the following extensive validation study as well as future applications on an even larger scale, we put an emphasis on performance improvements along the entire process chain and achieved a typical end-to-end runtime of a few hours.
An important factor in this regard is the straightforward and massive parallelization of the particle simulation, which allows for almost perfect scalability.

\bmhead{Overview of our validation study}
\begin{figure*}[htpb]%
  \centering
  \includegraphics[width=0.9\textwidth]{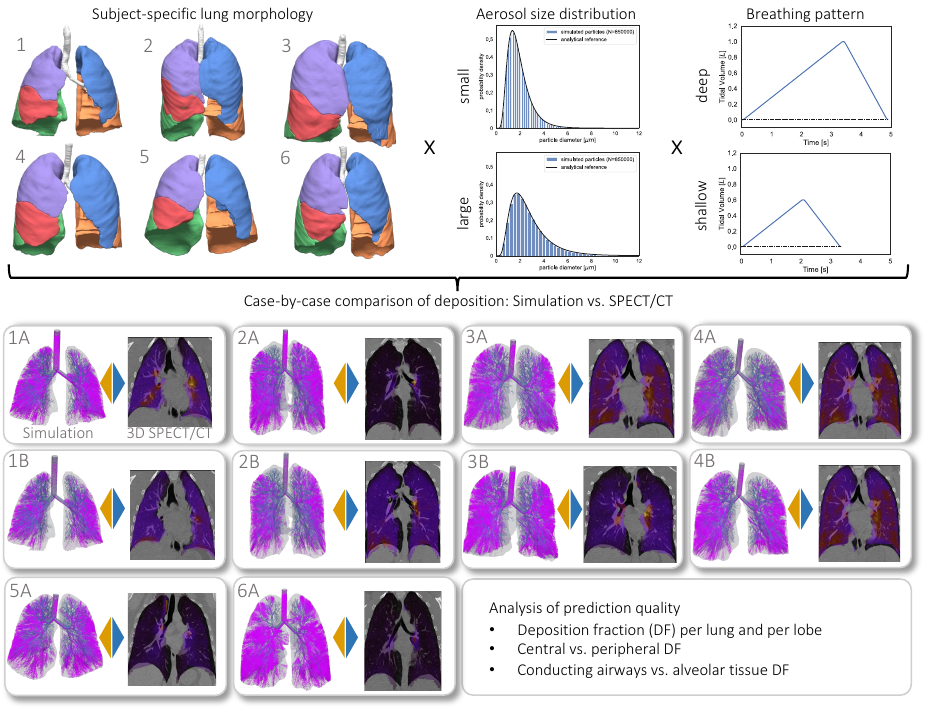}
  \caption{\textbf{Overview of our validation study design.}
      Six healthy human subjects, two different aerosol size distributions, and two different breathing patterns are combined in 10 inhalation experiments from the previously published, independent clinical study~\cite{conwayControlledParametricIndividualized2012} (see SI Sec.~\ref{sec:SI_exp_data} for a brief summary).
      Aerosol deposition patterns predicted by our simulations for these 10 inhalation experiments are compared to the in vivo reference data, mostly obtained from the 3D SPECT/CT images to analyze the model prediction quality.}
  \label{fig:study_design}
\end{figure*}%
Fig.~\ref{fig:study_design} provides an overview of the study design and validation results.
As a reference, we used the extensive experimental data set from a previous, independent clinical study, which has specifically been designed for the purpose of model validation~\cite{conwayControlledParametricIndividualized2012}.
This data set allows for the assessment of our model's prediction quality for a total of $10$ inhalation experiments by six healthy human subjects using one of two different aerosol sizes and breathing patterns.
Specifically the 3D SPECT/CT images taken after each inhalation experiment are used to validate the predicted aerosol deposition patterns as will be presented in the following sections.
See SI Sec.~\ref{sec:SI_model_methods}\ref{sec:SI_sim_setup_and_BCs} for details about the simulation setup and choice of boundary conditions to mimic the individual inhalation experiments from the original clinical study summarized in SI Sec.~\ref{sec:SI_exp_data}.

Before proceeding to the validation results, and in the light of a rapidly increasing number of data-driven approaches, it is worth emphasizing the physics- and physiology-based nature of the model.
None of the in vivo reference data was used to train our model as its predictive power stems solely from the fundamental laws of fluid and solid mechanics, and particle dynamics, in particular.
The model has further been developed based on physiological knowledge about anatomy and function of the human respiratory system.
It therefore offers the advantage to generalize without the need for large training data sets, which is naturally limited in this context due to ethical constraints and the limitation of measurement techniques.
As a consequence, the validation of this kind of physics- and knowledge-based model also requires much less reference data and comparable approaches have typically been validated studying one or two different human subjects.
To the best of our knowledge, this is the most comprehensive validation of an in silico model predicting aerosol deposition in the human lungs to date.

\bmhead{Validation of predicted lobar deposition}
\begin{figure*}[htpb]%
  \centering
  \includegraphics[width=0.9\textwidth]{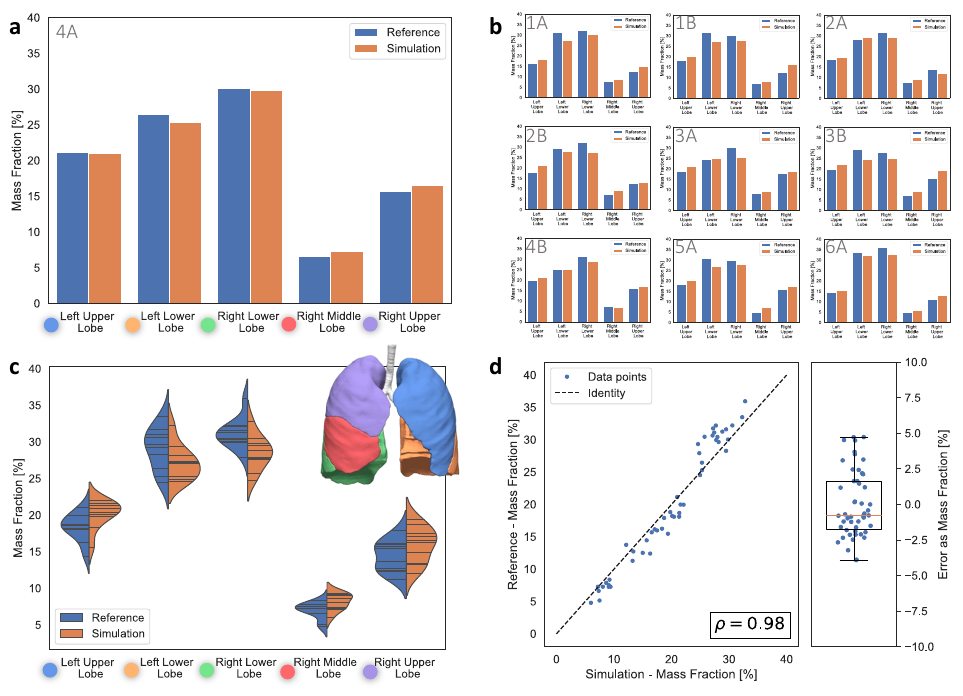}
  \caption{\textbf{Validation of predicted aerosol deposition per lobe.}
      \textbf{a} Deposited aerosol mass as a fraction of total aerosol mass deposited in the lungs. Comparison of model predictions (orange) and in vivo reference data (blue) for subject H04 (visit A).
      \textbf{b} Comparison of lobar deposition fractions for the remaining 9 inhalation experiments.
      \textbf{c} Distribution of model predictions (orange) and reference values (blue) for all 10 inhalation experiments (black lines indicate individual cases).
      \textbf{d} Scatter plot showing correlation and box plot showing error values for all 50 data points (10 inhalation experiments and 5 lobes), respectively.
          The error is defined as the mass fraction of deposited aerosol that is misattributed.}
  \label{fig:lobar_deposition}
\end{figure*}%
Fig.~\ref{fig:lobar_deposition} summarizes the validation results for the predicted aerosol deposition fraction per lobe.
Specifically, we consider the aerosol mass deposited in either of the five lobes as a fraction of the total aerosol mass deposited in the lungs.
Comparing the predictions from our simulation model (orange) to the in vivo reference data (blue) for an example subject H04 shows an excellent agreement (Fig.~\ref{fig:lobar_deposition}a) with less than $1 \%$ absolute difference for any lobe.
Naturally, the largest mass fraction deposits in the two larger lower lobes, followed by the two upper lobes and only about $7\%$ in the smallest right middle lobe.
This holds true if we look at the entire cohort of 10 inhalation experiments (Fig.~\ref{fig:lobar_deposition}a-b) and plot the distribution of model predictions (orange) and reference values (blue) in Fig.~\ref{fig:lobar_deposition}c.
The fact that the distribution of lobar deposition fractions agree very well for the entire cohort is a first important piece of evidence that our model is able to capture both general physiology and subject-specific factors in an adequate manner.
Thinking about future applications of our approach in clinical stages of drug development, where the results for an entire study population weigh much heavier than those of individual subjects, it is a particularly important finding that the model yields such good predictions from a population average point of view.

To quantify this overall prediction quality, we compare all 50 data points (10 inhalation experiments and 5 lobes) in the scatter plot in Fig.~\ref{fig:lobar_deposition}d and obtain an excellent correlation coefficient of $0.98$.
In addition, we show the distribution of errors in the box plot of Fig.~\ref{fig:lobar_deposition}d, which confirms the excellent prediction quality.
All error values lie within the range of $\pm 5\%$ meaning that less than $5\%$ of the mass is misattributed in any case.
For more than half of the predictions, the error is even smaller than $2\%$ as indicated by the interquartile range (black box).

Aside from the five lobes, we performed an identical analysis of the deposition fraction per left and right lung and present the results in SI Fig~\ref{fig:per_lung_deposition}, which reveals an overall even better prediction quality of the model with all error values in the range of $\pm 3\%$.

\bmhead{Validation of predicted central/peripheral deposition}
\begin{figure*}[htpb]%
  \centering
  \includegraphics[width=0.9\textwidth]{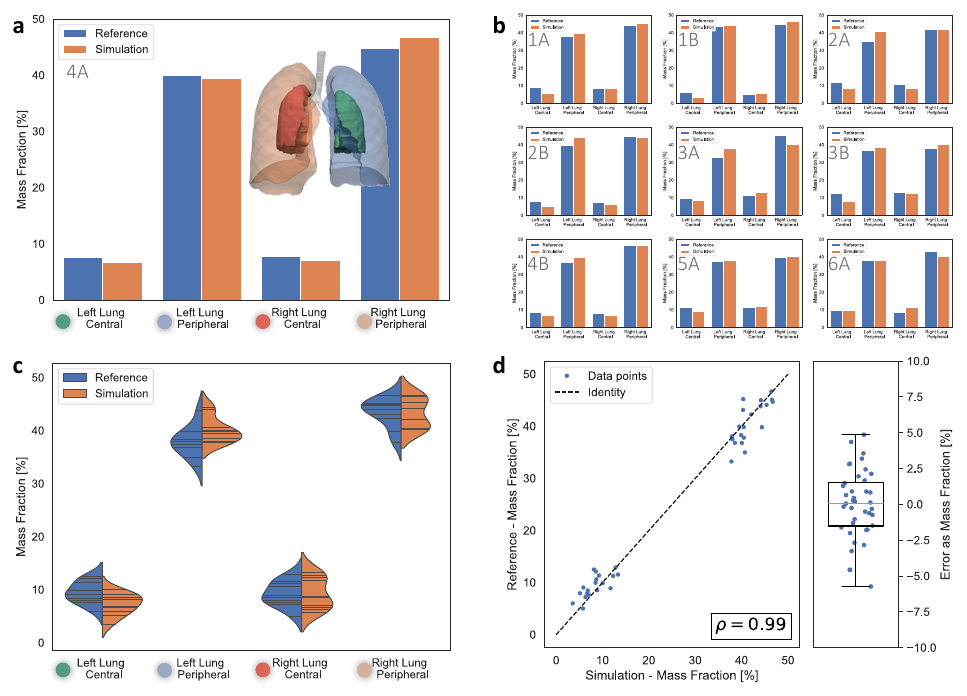}
  \caption{\textbf{Validation of predicted aerosol deposition in central versus peripheral regions.}
      \textbf{a} Deposited aerosol mass as a fraction of total aerosol mass deposited in the lungs. Comparison of model predictions (orange) and in vivo reference data (blue) for subject H04 (visit A).
      \textbf{b} Comparison of central/peripheral deposition fractions for the remaining 9 inhalation experiments.
      \textbf{c} Distribution of model predictions (orange) and reference values (blue) for all 10 inhalation experiments (black lines indicate individual cases).
      \textbf{d} Scatter plot showing correlation and box plot showing error values for all 40 data points (10 inhalation experiments and 4 regions), respectively.
          The error is defined as the mass fraction of deposited aerosol that is misattributed.}
  \label{fig:cp_deposition}
\end{figure*}%
Fig.~\ref{fig:cp_deposition} shows the validation results for the central versus peripheral deposition in both lungs.
The spatial subdivision into central and peripheral parts of each lung is a widely used proxy for the actual question how much aerosol deposits in the (large) conducting airways as compared to the alveolar tissue (and the small, distal airways).
Even though it is known that these central regions also include considerable amounts of alveolar tissue and that the peripheral regions also contain conducting airways, this analysis is the closest one can get with the limited spatial resolution of in vivo measurement techniques.
Although our model allows for a much finer spatial resolution of deposition results and thus more direct insights regarding these questions (see the following sections), we decided to still validate the central and peripheral deposition predictions to reflect the state of the art.
We classified the central and peripheral regions in each lung based on concentric shells around the hilum as described in SI Sec.~\ref{sec:SI_post_processing}\ref{sec:SI_c2p_split_analysis} and illustrated in the inset of Fig.~\ref{fig:cp_deposition}a.
Apart from these four regions of interest (left-lung-central, left-lung-peripheral, right-lung-central, right-lung-peripheral), the following analysis is analogous to the preceding validation of lobar deposition.
Fig.~\ref{fig:cp_deposition}a again shows the results for one example of subject H04.
As expected, the majority of aerosol mass deposits in the peripheral parts, because they are larger by volume and the aerosol size in this example is fine enough to reach these peripheral parts of the lung.
More importantly, excellent agreement between simulation results and in vivo reference data is obtained for all four regions with a maximum absolute difference of about $2\%$.
The same comparison for the remaining 9 inhalation experiments is shown in Fig.~\ref{fig:cp_deposition}b for completeness.
Looking at the distributions for the entire study cohort in Fig.~\ref{fig:cp_deposition}c again reveals a very good overall prediction quality and confirms the correctness of the model also on a population-average level.

Quantitative comparison of simulation and experiment for all 40 data points (10 inhalation experiments, 4 regions) yields the scatter plot shown in Fig.~\ref{fig:cp_deposition}d and a correlation coefficient of $0.99$.
Here, the corresponding errors reside in between $\pm 6\%$ and are thus only slightly wider spread than for the previously considered lobar deposition.
For more than half of the predictions, the error is once again smaller than $2\%$ as indicated by the interquartile range (black box).
Keeping in mind also the uncertainties and limitations of the in vivo reference data, this should again be considered an excellent result for a physics- and physiology-based model.

\begin{figure*}[htpb]%
  \centering
  \includegraphics[width=0.9\textwidth]{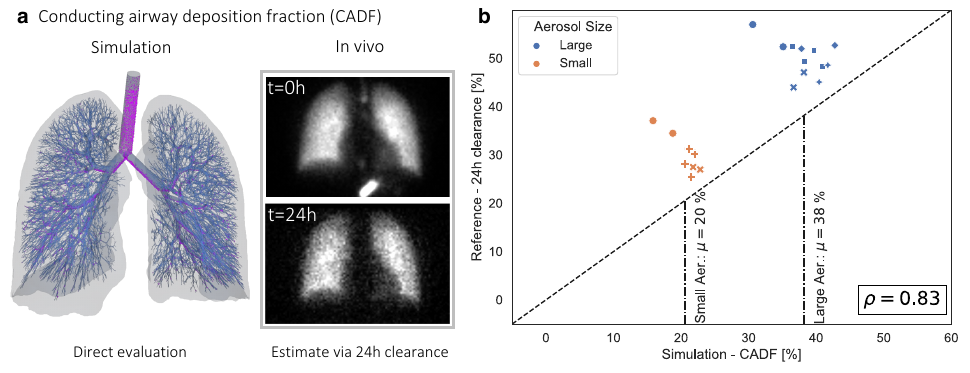}
  \caption{\textbf{Validation of predicted aerosol deposition fraction in conducting airways versus alveolar tissue.}
      \textbf{a} Illustration of measured 24h clearance and conducting airway deposition fraction (CADF) obtained from simulation results.
      \textbf{b} Scatter plot of measured 24h clearance versus predicted CADF for all 20 data points (10 inhalation experiments, left and right lung).
          Color indicates small (orange) and large (blue) aerosol particle size and marker type indicates different human subjects.}
  \label{fig:cadf}
\end{figure*}

\bmhead{Validation of predicted deposition in conducting airways versus alveolar tissue and the influence of aerosol particle size}
One of the crucial questions of targeted pulmonary drug delivery is how much aerosol deposits in the (large) conducting airways as compared to the respiratory zone and alveolar tissue.
Depending on the disease and the drug's intended site(s) of action, the target may be different, but the ability to predict the dose at the local site of action is an invaluable asset and opens the door to an optimization of aerosol and inhalation parameters.
Due to its unprecedented level of comprehensiveness and spatiotemporal resolution, our model is the first to address this unmet need.
The validation of conducting airway deposition fraction (CADF) is challenging, however, because it cannot be measured directly in a clinical setting with human subjects.
We therefore fall back to 24h clearance data obtained from the comparison of two gamma scintigraphy images in the reference data set as illustrated in Fig.~\ref{fig:cadf}a.
Based on the design of the original clinical study, this was considered a measure of deposition in the conducting airways due to the substantial difference in clearance rates in conducting airways as opposed to the alveolar tissue~\cite{majoralControlledParametricIndividualized2014,ilowiteRelationshipTracheobronchialParticle1989}.
Fig.~\ref{fig:cadf}b shows the correlation of measured 24h clearance and predicted CADF for all 20 data points (10 inhalation experiments, left and right lung).
Although one cannot expect a 1:1 equivalence of these two measures, the obtained correlation coefficient of $0.83$ indicates a very good prediction quality of the model also in this regard.
The most remarkable aspect of these results, however, are the two clearly distinguishable clusters of data points from those experiments with small (orange) and large aerosol size (blue), respectively.
As expected, smaller aerosol size leads to smaller CADF and in turn more deposition in the alveolar tissue and vice versa.
The ability of the model to correctly predict the influence of the particle size on the aerosol deposition pattern is yet another important piece of evidence for the validity and in turn utility of the model.

\bmhead{Total mass balance and high-resolution deposition analysis}
\begin{figure*}[htb]%
  \centering
  \includegraphics[width=0.6\textwidth]{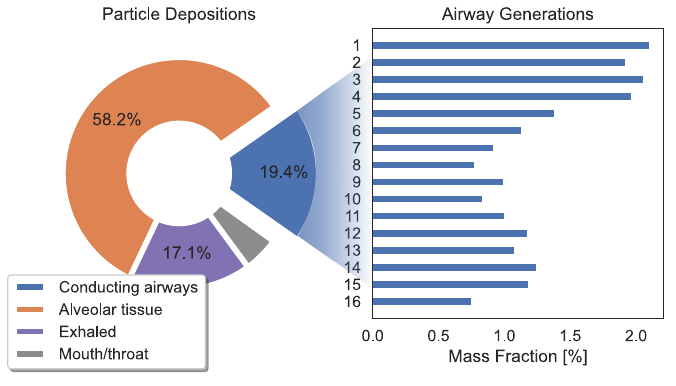}
  \caption{\textbf{Example for detailed aerosol deposition analysis for case 4A.}
      Left: Total mass balance with mass fractions of inhaled aerosol.
      Note that aerosol particles in motion (\textless 1\%) are left out for clarity.
      Right: Mass fraction of deposited aerosol per airway generation from trachea (generation number 1) to smallest conducting airways (generation number 16).}
  \label{fig:mass_balance_and_dep_per_airway_gen}
\end{figure*}%
Because we include all relevant components of the human respiratory tract into the model, and because our approach is resource-efficient to achieve high spatiotemporal resolution, our results allow novel analyses that are not possible with state-of-the-art subject-specific models.
For instance, the total mass balance shown in Fig.~\ref{fig:mass_balance_and_dep_per_airway_gen} requires a physically and physiologically complete model, i.e., the inclusion of all major functional units in their entirety \textit{and} the capability to simulate complete breathing patterns (inhalation and exhalation) of aerosol in this closed system.
Moreover, the high spatiotemporal resolution allows for in-depth analyses of, e.g., the deposition fraction per airway generation shown in Fig.~\ref{fig:mass_balance_and_dep_per_airway_gen} and simultaneously down to the individual particle level as illustrated in Fig.~\ref{fig:model}d.
These are truly distinguishing features of our novel approach, which not only outperforms existing in silico approaches, but also surpasses the capabilities of in vivo methods such as nuclear imaging.

\bmhead{Tissue disease modeling}
This section briefly showcases our unique ability to account for diseases affecting the alveolar tissue and small airways. 
Disease-affected lungs are characterized by a heterogeneous distribution of pathological changes, which in turn results in, e.g., heterogeneous distribution of tissue stiffness or fluid accumulations within lung sections.
From a modeling perspective, it is therefore crucial to include the alveolar tissue throughout the entire lung and take possible spatial differences of its mechanical properties into account.
Being able to directly assign the pathologically altered material behavior to the alveolar cluster elements in a consistent and spatially highly resolved manner is a clear advantage over existing modeling approaches, which try to approximate the effect of disease by means of modified boundary conditions applied to the truncated lung models.
In this work, we consider a patient suffering from idiopathic pulmonary fibrosis (IPF) as a showcase example to demonstrate our model's capabilities.
\begin{figure*}[htb]%
  \centering
  \includegraphics[width=0.95\textwidth]{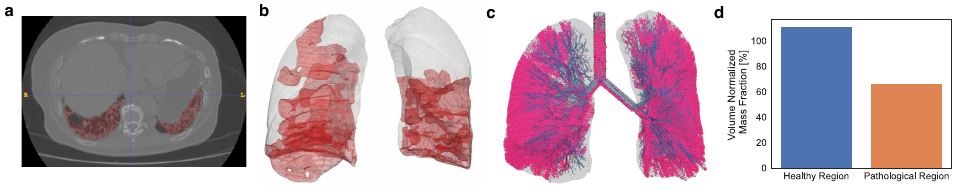}
  \caption{\textbf{Modeling pulmonary diseases such as fibrosis.}
      \textbf{a} Annotation of fibrotic regions in the CT image (red color).
      \textbf{b} Adaptation of the computational model by assigning a higher stiffness to fibrotic regions (red color).
      \textbf{c} Deposition pattern resulting from simulation.
      \textbf{d}~Quantitative comparison of volume-normalized aerosol deposition fraction in healthy versus pathological regions.}
  \label{fig:disease_modeling}
\end{figure*}%
The patient's lung is shown in Fig.~\ref{fig:disease_modeling}.
IPF affected regions showing honeycombing or reticulation in the CT scan were annotated and assigned a significantly higher stiffness to model the known stiffening of fibrotic scar tissue (marked red in Fig.~\ref{fig:disease_modeling}a and b).
Quantitative analysis of the differences in the resulting aerosol deposition pattern (Fig.~\ref{fig:disease_modeling}c) between healthy and pathological regions shows that considerably less aerosol deposits in the pathological regions.
For this specific example, the volume-normalized fraction of deposited aerosol mass was reduced by about $40\%$ as shown in Fig.~\ref{fig:disease_modeling}d.
This highlights the relevance for informed dosing decisions in the treatment of lung diseases via locally acting, orally inhaled drugs. 
At this point we would like to emphasize that to date no other in silico approach can provide this quantitative information about the local effective dose and nuclear imaging is currently the only means to approximate this quantity in disease-affected lungs.
However, even leaving ethical concerns aside, it would be hard if not impossible to incorporate SPECT imaging into large clinical trials for certain diseases to obtain quantitative information about the local effective dose at the site of action.
The same unmet need applies to bioequivalence studies of generic drug products, which need to show equivalence of the locally delivered dose with the reference product to get regulatory approval for their drug-device combination.

\section{Discussion}\label{sec:discussion}
We have presented and validated a novel physics-based in silico approach that enables the subject-specific non-invasive prediction of aerosol deposition throughout the entire human lung with excellent accuracy.
Its unprecedented spatiotemporal resolution allows to track each aerosol particle anytime during the breathing cycle and, for the first time, anywhere in the entire conducting airway tree as well as the alveolar region.
Rigorous validation against 3D SPECT/CT images from a healthy human cohort ($N=10$ inhalation experiments) shows excellent agreement with all errors within $\pm 5 \%$ for lobar and $\pm 6 \%$ for central/peripheral deposition fractions.
For more than half of the predictions, the error is even smaller than $2 \%$.
In addition, the nicely matching distributions of predicted mass fractions over the entire study cohort demonstrate the correctness of this physics- and physiology-based model on a population-average level.
The newly accessible deposition fraction in the entire tree of conducting airways (CADF) is validated by means of 24h clearance data and shows a high correlation ($\rho=0.83$) with this surrogate reference data.
Moreover, the known strong influence of the aerosol particle size on CADF, and inversely the alveolar tissue deposition fraction, is clearly reflected in the model predictions and exemplifies the vast opportunities to analyze all kinds of influences such as aerosol, breathing or subject-specific parameters. 
We further showcased the model's capabilities to account for strong tissue and airflow heterogeneities in diseased lungs by studying an IPF patient's lung.

The key findings of this paper lay the groundwork for a number of high-impact use cases in healthcare and life sciences.
The proven accuracy of our approach enables for the first time the prediction and optimization of the local effective dose in arbitrary (spatial or generational) subdomains of both healthy and diseased lungs.
Hence, our technology can be used to demonstrate bioequivalence between an existing and a new generic drug/device combination, thereby reducing the costs  for generics and making them more widely available in the long run.
In addition, our technology provides the first cost-effective means of optimizing pulmonary drug delivery with respect to delivery device and aerosol parameters. 
Furthermore, the unique combination of quantitative imaging, in silico functional analysis of the lung allows us to extract unprecedented levels of information from a subject's lung which can be condensed into digital biomarkers that are much more sensitive to subtle changes in lung function, e.g., due to medication, than current pulmonary function tests. Our technology can thus be levered to overcome the challenge of dependence on fuzzy clinical endpoints and limited sensitivity of current pulmonary function test in the development of respiratory therapies through in silico analysis of lung function. In combination with high-resolution deposition statistics, these biomarkers can not only be used to inform and de-risk all stages of pre clinical drug development, but also help to define personalized treatment protocols paving the way for companion diagnostic in respiratory care by means of accompanying predictive in silico modeling. 

Beyond that we envision our approach to be an effective and cost-efficient alternative to nuclear imaging of the lung due to its far better risk-to-benefit ratio as compared to, e.g., SPECT/CT or PET/CT.
Featuring a much higher spatiotemporal resolution and no radiation exposure, our physics-augmented "imaging modality" can lead to more precise personalized diagnoses for a significantly broader patient population.

Lastly, we would like to stress that the high levels of computational efficiency and automation in model generation and calibration we demonstrated in this validation study render the applications sketched above not a distant dream but a realistic possibility in the near future since our technology is ready to be deployed at scale.
We also plan to combine this with our previous work \cite{brandstaeterGlobalSensitivityAnalysis2021a, wirthlGlobalSensitivityAnalysis2023} and add an optimization and sensitivity analysis layer to our computational pipeline.
This will not only pave the way for optimization of device and aerosol parameters, but will also allow us to identify beneficial and detrimental physiological or geometric characteristics of patients for instance to inform stratification approaches in clinical trials.

\section{Methods}\label{sec:methods}

\bmhead{In vivo reference data}
We used the in vivo reference data collected in an independent clinical study previously published in~\cite{conwayControlledParametricIndividualized2012,majoralControlledParametricIndividualized2014}.
It was designed specifically for the purpose of model validation and therefore provides an unparalleled quality of data describing the aerosol deposition patterns by means of 3D SPECT/CT and, among others, 2D gamma scintigraphy images, in combination with precisely controlled and measured inputs such as aerosol size distribution and breathing pattern.
The data set includes six healthy human subjects with a total of 10 different inhalation experiments used for the presented validation study.
See SI Sec.~\ref{sec:SI_exp_data} for more details.

\bmhead{Subject-specific model generation}
For each of the study participants, we generated a subject-specific computational model starting from the available CT image.
Beside the volumetric representation of airways, lungs and lobes, we also made use of the spirometry data to determine the material parameters of our whole lung model in a subject-specific manner.
These steps are described in more detail in SI Sec.~\ref{sec:SI_model_methods}\ref{sec:SI_model_gen} and \ref{sec:SI_model_methods}\ref{sec:SI_material_param_determination}, respectively.
To ensure reproducibility and scalability, the entire model generation process has been automated by implementing it as in-house software packages and integrating it in an end-to-end computing pipeline.

\bmhead{Simulation setup and boundary conditions}
The simulation setup and choice of boundary conditions mimic the original inhalation experiments from the in vivo reference data set summarized in SI Sec.~\ref{sec:SI_exp_data}.
This mainly involves the transient flow of the defined breathing patterns and the particle size distribution as well as other aerosol specifications.
Refer to SI Sec.~\ref{sec:SI_model_methods}\ref{sec:SI_sim_setup_and_BCs} for the detailed setup.

\bmhead{Simulation software}
The entire fluid/structure and particle simulation software is implemented as professionally maintained in-house C++ libraries and Python packages and does not rely on any closed-source software products of third parties.
Instead, we build upon well-established open-source packages including Trilinos~\cite{thetrilinosprojectteamTrilinosProjectWebsite2020}, numpy~\cite{harrisArrayProgrammingNumPy2020}, and scipy~\cite{virtanenSciPyFundamentalAlgorithms2020}.
This not only enabled the new, innovative model design combining a 3D Lagrangian particle transport and deposition model with the reduced-dimensional whole lung model for airflow and local tissue deformation, but also opens the door for future extension and customization.
Moreover, our focus on high-performance software and computing environments is an important factor for large-scale in silico trials.

\bmhead{Post-processing and data analysis}
Aerosol deposition statistics were calculated from the raw simulation output, which mainly consists of the location and state of every individual particle over time.
From the final particle deposition pattern, we computed the aerosol mass (fraction) within each spatial region of interest such as the lungs, lobes, or the central/peripheral parts.
The same was done for the SPECT/CT image data of the experiments to then compare the results and compute the correlation and error measures.
In addition, we analyzed the deposited mass from simulations based on functional categories such as alveolar tissue, conducting airways, mouth/throat or exhaled.
These steps are described in more detail in SI Sec.~\ref{sec:SI_post_processing}.
All post-processing and data analysis steps were implemented as in-house Python packages.
The visualizations of simulation results and reference data were created using vtk~\cite{schroederVisualizationToolkit4th2006}, Paraview~\cite{ahrensParaViewEndUserTool2005}, and ITK-SNAP~\cite{yushkevichUserguided3DActive2006}.

\backmatter

\bmhead{Supplementary information (SI)}
Supplementary information is provided in a separate file attached to this article. 

\bmhead{Acknowledgments}
We are grateful to Ira Katz for the provision of the in vivo data set from \cite{conwayControlledParametricIndividualized2012,majoralControlledParametricIndividualized2014}.
W.A.W. gratefully acknowledges the financial support by BREATHE, a Horizon 2020-ERC-2020-ADG project (101021526-BREATHE).

\bmhead{Competing interests}
J.Bi., K.R.W., K.W.M., D.R., and M.J.G.~are inventors on a pending patent application.
All other authors declare no competing interests.

\bmhead{Data availability}
All the data described and generated in the paper and SI is available upon reasonable request from the corresponding author. 

\bmhead{Code availability}
All results described in this paper were created using our own in-house codebase.
To protect this asset, we are unfortunately not able to share the source code at this point in time. 

\bmhead{Authors' contributions}
M.J.G., J.Bi., W.A.W., and K.W.M. designed the research.
K.R.W., M.J.G., and D.R. developed the computational model.
D.R., J.R., M.J.G., M.B., K.R.W., M.R., J.B{\"u}., N.P., and J.Bi. contributed to software development.
M.J.G., K.R.W., and M.R. performed the simulation setup and runs.
J.R., M.R., M.J.G, and K.W.M. did post-processing, data analysis and visualization of simulation results.
M.B. and N.P. analyzed the in vivo reference data.
M.J.G., J.Bi., K.R.W, and K.W.M. wrote the manuscript.
All authors reviewed the manuscript.\\

\bibliography{ebenbuild}

\end{document}

% work around broken cross-references between main article and SI document
\makeatletter\@input{main_article_xr_references.tex}\makeatother

\maketitle

\SItext

\section{Additional information on in vivo reference data}
\label{sec:SI_exp_data}
We use the in vivo reference data collected in an independent previous clinical study and published in \cite{conwayControlledParametricIndividualized2012}.
This study has been designed specifically for the purpose of model validation and provides an unprecedented amount and quality of both imaging and other clinical data related to aerosol inhalation and deposition in the respiratory tract of human subjects.
The most important point is that the experiments were parametrized and controlled in a manner that only one of the three defined main influences (particle size, depth of breathing, and carrier gas) were changed at a time.
The main study, which is used as the source of reference data in this article, included six healthy male subjects with two inhalation experiments each, i.e., a total of 12 cases.
We excluded the two cases using helium as carrier gas, because this is beyond the scope of this article, and used the remaining 10 cases for our model validation.
For each of the cases, the resulting aerosol deposition was measured by 3D SPECT/CT imaging after inhalation of a radiolabeled aerosol with defined and measured particle size distribution and defined and enforced breathing pattern.
Specifically, a constant inspiratory flow rate of \qty{18}{\liter/\minute} was applied over the fixed inhalation time of either \qty{2.0} or \qty{3.33}{\second}, thus resulting in either \qty{600} or \qty{1000}{\milli\liter} of tidal volume for the shallow and deep breathing pattern, respectively.
The measured aerosol properties varied slightly over the cases, but on average the mass median aerodynamic diameter (MMAD) was \qty{3.1} and \qty{5.7}{\micro\meter} for the small and large aerosol, respectively.
An overview of the cases used in the present article is provided in Table~\ref{tbl:cases}.
To be able to extract the geometry of the airway tree, lobes and lungs, an HRCT image of each patient has been taken on a different occasion before the actual inhalation experiments.
Further measurements such as 24h clearance fraction determined as the difference of two planar gamma scintigraphy images and other patient data such as height, weight, age, and spirometry data complete the supplementary data set published together with the original article~\cite{conwayControlledParametricIndividualized2012}.
An in-depth analysis of this experimental in-vivo data and a preliminary comparison to simulation has been published in follow-up articles~\cite{katzControlledParametricIndividualized2013,majoralControlledParametricIndividualized2014} and the study has later been extended to asthmatics in~\cite{flemingControlledParametricIndividualized2015}.
\begin{table}[hbt]
  \centering
  \begin{tabular}{l|c|c|c|c}
    Case ID & Subject & Visit & Aerosol size & Breathing pattern\\
    \hline
    1A & H01 & A & Large & Shallow \\
    1B & H01 & B & Small & Shallow \\
    \hline
    2A & H02 & A & Large & Deep \\
    2B & H02 & B & Small & Deep \\
    \hline
    3A & H03 & A & Large & Shallow \\
    3B & H03 & B & Large & Deep \\
    \hline
    4A & H04 & A & Small & Shallow \\
    4B & H04 & B & Small & Deep \\
    \hline
    5A & H05 & A & Large & Shallow \\
    \hline
    6A & H06 & A & Large & Deep
  \end{tabular}
  \vspace{1em}
  \caption{Overview of cases from \cite{conwayControlledParametricIndividualized2012} used for the validation study in the present article.}
  \label{tbl:cases}
\end{table}

\section{Additional information on computational model and methods}
\label{sec:SI_model_methods}

\subsection{Patient-specific model generation}
\label{sec:SI_model_gen}
Patient-specific computational models of the six healthy subjects were created according to the steps described in the following and based on the data set described in \cite{conwayControlledParametricIndividualized2012} and summarized above.
The geometry of the patient's lungs, including lobes, initial parts of the airway tree and its centerline, is extracted from the CT imagery as shown in the model creation overview in Fig.~\ref{fig:model} of the main article.
While we usually employ computer vision and deep learning techniques to automatically segment the CT images, we utilized the segmentation masks that are described in \cite{conwayControlledParametricIndividualized2012} and were kindly shared with us by the authors of the study.
Due to resolution limitations, it is generally not possible to extract the whole airway tree from the CT scan.
The higher generation airways are thus generated using a recursive space-filling tree growth algorithm \cite{ismailCoupledReducedDimensional2013,tawhaiGenerationAnatomicallyBased2000}, which results in hybrid patient-specific/morphometric airway trees as shown in Fig.~\ref{fig:model}b.
This highly patient-specific geometry of the lungs and the airway tree is then translated into a physics-based simulation model that accounts for, among others, transient airflow in the airways and alveoli including elastic interaction with
the rib cage and diaphragm (see SI Sec.~\ref{sec:SI_model_methods}\ref{sec:SI_lungmodelflow} for details).
Likewise, this patient-specific geometry is used in the newly developed model for particle transport and deposition in the whole lung (see SI Sec.~\ref{sec:SI_model_methods}\ref{sec:SI_particletransport} for details).
Finally, the material properties of conducting airways, alveolar tissue, and chest cavity are calibrated using information from the imaging and functional data from the experimental data set \cite{conwayControlledParametricIndividualized2012} and population averages.
Details about this process are provided in SI Sec.~\ref{sec:SI_model_methods}\ref{sec:SI_material_param_determination}.

\subsection{Computational whole lung model for airflow and local deformation of lung tissue}
\label{sec:SI_lungmodelflow}
Previous work has shown that this modeling approach can accurately predict airflow and local deformation in both spontaneous breathing as well as mechanical ventilation~\cite{ismailCoupledReducedDimensional2013,rothComprehensiveComputationalHuman2017, rothComputationalModellingRespiratory2017}.
Temporal and regional differences in ventilation, often referred to as asynchrony and asymmetry, are naturally accounted for by the model, as these arise from the physics-based equations that are solved.
While these effects might be negligible for normal breathing in healthy patients, these effects are relevant in patients with respiratory diseases.
Even in these scenarios it has been shown that the applied model is able to predict the behavior of lungs~\cite{rothCouplingEITComputational2017}.

\subsubsection{Conducting airways}
\begin{figure}[htb]
  \begin{center}

      \ctikzset{bipoles/thickness=2}
      
      \begin{circuitikz}[scale=0.6]
		\usetikzlibrary{calc}

		\node at (-13.7,8.5) {\textbf{a}};
		\node at (-9,3.5) {\includegraphics[height=6cm]{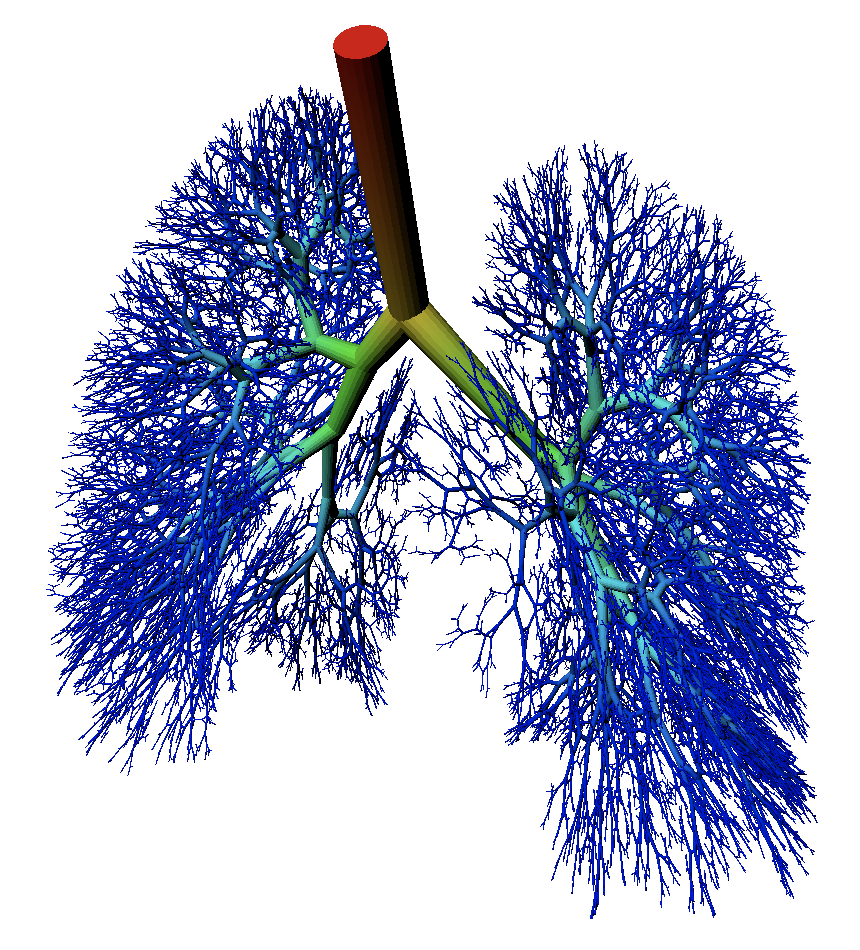}};

		\node at (-2.5,8.5) {\textbf{b}};
		\draw [thick] (1.2,6.5) -- (0.8,0) -- (-0.8,0) -- (-1.2,6.5) -- (1.2,6.5);
		\draw [dashed, black!50] (0,6.5) -- (0,0);
		\draw [fill=black] (0,6.5) circle (.7ex);
		\draw (0,6.5) node[anchor=south] {$(P_1,Q_1)$};
		\draw[fill=black] (0,0) circle (.7ex);
		\draw (0, 0) node[anchor=north] {$(P_2,Q_2)$};

		\draw (-1,4) arc(180:360:1 and 1.6);
		\draw (-1,4) to (1,4);
		\draw [-latex] (0.0,4) to (0.0,2.4);
		\draw [-latex] (0.33,4) to (0.33,2.5);
		\draw [-latex] (0.66,4) to (0.66,2.82);
		\draw [-latex] (-0.33,4) to (-0.33,2.5);
		\draw [-latex] (-0.66,4) to (-0.66,2.82);
		
		\draw (1,3.5) node[right] {$v = v(x,t)$};

	    \draw [-latex] (-1.5,1) -- (-0.9,1);
	    \draw [latex-] (0.9,1) -- (2.5,1) node[anchor=south] {$A=A(x,t)$};
	    \draw (-1.5,1) -- (2.5,1);

		\node at (6,8.5) {\textbf{c}};
		\draw [-latex] (8,7.5) node[anchor=south] {$Q_1$} -- (8,6.7);
        \draw (8,6.5) node[left]{$P_1$}
          to [short, *-] (8,6)
          to [R, l=$R_{conv}$, *-] (8,4)
          to [R, l=$R_{\mu}$, *-*] (8,2)
          to [L, l=$I$, -*] (8,0)
          node[left]{$P_2$}
          (8,6) to [C, l=$C$, *-] (11,6)
          to [R, l=$R_{visc}$, *-*] (13,6)
          node[right]{$P_{ext}$};
      \draw [-latex] (8,0) -- (8,-0.8) node[anchor=north] {$Q_2$};
      
      \end{circuitikz}

  \end{center}
  \caption{\textbf{Conducting airway model.}
           \textbf{a} Example for the full, 16 generation airway tree used for subject H04 in the present work (approx.~35,000 airway elements, colored by diameter).
           \textbf{b} Schematic representation of a single airway element including the flow related quantities in this reduced-dimensional formulation.
           \textbf{c} Illustration of the model components using the analogy of an electric circuit.
               One-sided version of the symmetric airway model is used here for brevity.
               Image adapted from~\cite{ismailCoupledReducedDimensional2013}.}
  \label{fig:red_dim_airway_model}
\end{figure}

The computation of the transient airflow in the conducting airway tree is based on a very efficient, yet accurate reduced-dimensional formulation developed in \cite{ismailCoupledReducedDimensional2013,alastrueyLumpedParameterOutflow2008,rothComprehensiveComputationalHuman2017, rothComputationalModellingRespiratory2017} and illustrated in Fig.~\ref{fig:red_dim_airway_model}.
The dimensional reduction of the problem is achieved by integrating the Navier-Stokes equations over the domain, exploiting information about the geometry of the airways as well as the flow within the airways.
Specifically, it is assumed that airways have axisymmetric cross-sections and negligible curvature and that the flow in the airways likewise has an axisymmetric velocity profile.
The reduced-dimensional formulation can then be obtained by first integrating along the radial direction of the airways and subsequently along the axial direction of the airways. 
For further details on the derivation of the reduced-dimensional problem, the reader is referred to, e.g., \cite{ismailCoupledReducedDimensional2013,alastrueyLumpedParameterOutflow2008}.
While the originally published model takes into account elasticity of the airway walls, we found that its influence on the particle deposition results of the present study is negligible and therefore used non-compliant airways for simplicity.

For brevity a one-sided version of the symmetric airway model is schematically depicted in Fig.~\ref{fig:red_dim_airway_model}c and the corresponding equations relating airway pressure~$P$ and flow rate~$Q$ are given as:
\begin{equation*}
  \begin{split}
    C \frac{dP_{\mathrm{1}}}{dt} + Q_{\mathrm{2}} - Q_{\mathrm{1}}  &= 0 \\
    I\frac{dQ_{\mathrm{2}}}{dt} + R Q_{\mathrm{2}} +P_{\mathrm{2}} -P_{\mathrm{1}} &= 0.
  \end{split}
\end{equation*}
In analogy to components of an electric circuit, $C$ denotes the capacitive, $I$ the inductive, and $R$ the resistive part of the system, where only $R$ remains active for non-compliant airways.
The pressure and flow rate at the in- and outlets are denoted by $P_{\mathrm{1}}$, $P_{\mathrm{2}}$ and $Q_{\mathrm{1}}$, $Q_{\mathrm{2}}$, respectively. 
The capacitive part represents the effects of airway compliance and the inductive term models inertial effects of both the air and the airway wall.
Finally, the resistive term is used to model dissipative effects that occur due to the viscosity of air.
A suitable nonlinear resistance model that takes into account turbulent as well as geometric losses within the airway tree, has been developed in \cite{pedleyPredictionPressureDrop1970} and refined in \cite{vanertbruggenAnatomicallyBasedThreedimensional2005} and is also used in our model.

\subsubsection{Alveolar clusters}
\begin{figure}[htb]
  \begin{center}
    \textbf{a}
    \begin{minipage}[t][5cm][c]{0.35\textwidth}%
      \centering
      \includegraphics[width=0.9\textwidth]{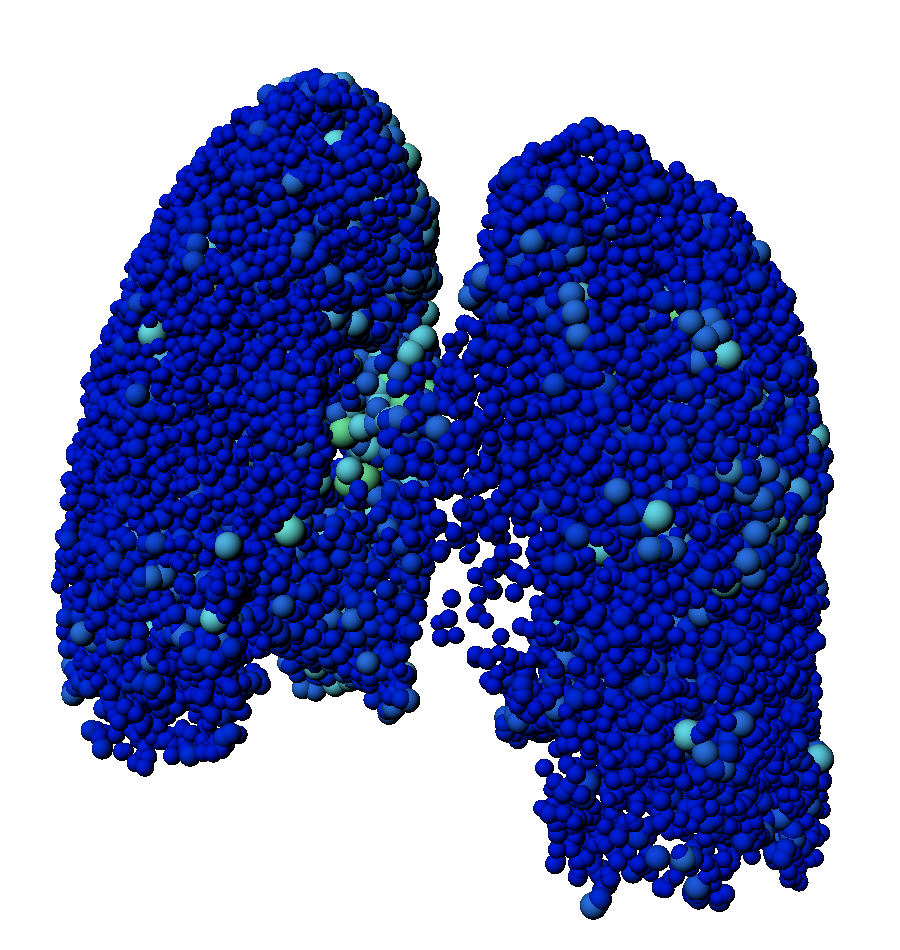}
    \end{minipage}%
    \hspace{2em}
    \textbf{b}
    \begin{minipage}[t][5cm][c]{0.2\textwidth}%
      \centering
      alveolar duct:\\\vspace{2mm}
      \includegraphics[width=0.6\textwidth]{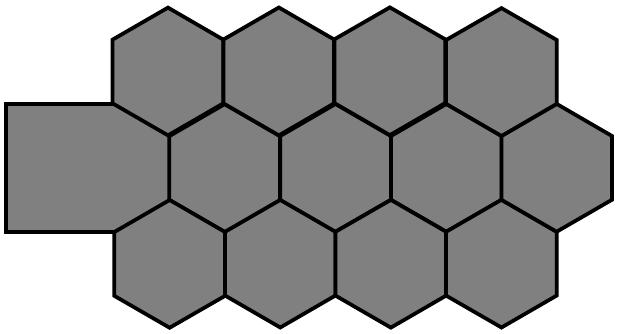}\\\vspace{3mm}
      alveolar cluster:\\\vspace{2mm}
      \includegraphics[width=0.7\textwidth]{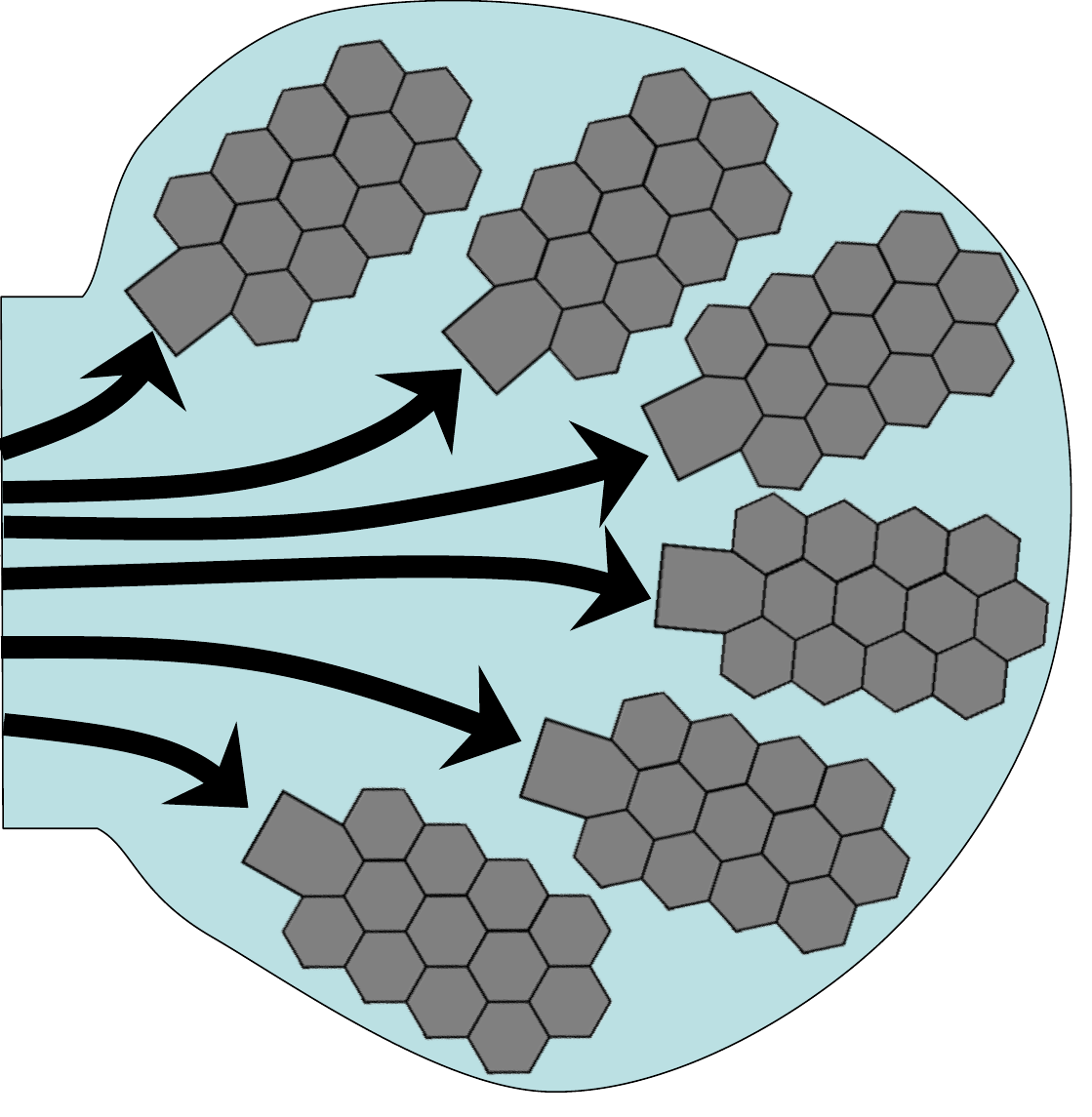}
    \end{minipage}%
    \hspace{2em}
    \textbf{c}
    \begin{minipage}[t][5cm][c]{0.3\textwidth}%
      \centering
      \begin{circuitikz}
	    \draw [-latex] (0.5,0) node[left] {$Q_{i}$} -- (1.0,0);
	    \draw (2.5, 0) node {$P_\mathrm{a}$};
	    \draw (4.5, -0.25) node {$P_\mathrm{pl/intr}$};
	    \draw (1.0, 0.25) to [short, -] (1.25, 0.25);
	    \draw (1.0, -0.25) to [short, -] (1.25, -0.25);
	    \draw (1.25, 0.25) to [short, -] (1.25, 3.5);
	    \draw (1.25, -0.25) to [short, -] (1.25, -0.5);
	    \draw (1.25, 0.5) to [short, -] (4.0, 0.5);
	    \draw (1.25, -0.5) to [short, -] (4.0, -0.5);
	    \draw (3.75, 0.45) to [short, -] (3.75, -0.45);
	    \draw (3.75, 0) to [short, -] (4.25, 0);
	    \draw (4.25, 0) to [short, -] (4.25, 3.5);
	    \draw (1.25, 3.25) to [damper, l=$B_\mathrm{a}$, -] (4.25, 3.25);
	    \draw (1.25, 2.25) to [R, l=$E_1$, -] (4.25, 2.25);
	    \draw (1.25, 1.25) to [R, l=$E_2$, -] (2.75, 1.25);
	    \draw (2.75, 1.25) to [damper, l=$B$, -] (4.25, 1.25);
      \end{circuitikz}
    \end{minipage}%
  \end{center}
  \caption{\textbf{Alveolar cluster model.}
           \textbf{a} Illustrative example of the full cloud of approx.~18,000 alveolar cluster elements used for subject H04 in the present work.
               Each alveolar cluster is depicted as a sphere with a volume equivalent to the gas volume extracted from the CT image and is colored by that volume.
           \textbf{b} Schematic of an alveolar duct and a grouping of them forming an alveolar cluster.
               Image adapted from~\cite{ismailCoupledReducedDimensional2013}.
           \textbf{c} Four element Maxwell model of an alveolar duct, with acinar pressure $P_{\mathrm{a}}$, pleural/inter-acinar pressure $P_{\mathrm{pl/intr}}$, and flow rate $Q_i$.
           }
  \label{fig:alveolar_cluster_model}
\end{figure}

The respiratory zone beyond the distal ends of the conducting airway tree consisting of respiratory bronchioles and alveolar tissue is represented by a further reduced-dimensional model component we refer to as \textit{alveolar cluster}.
An example for the resulting cloud of approx.~18,000 alveolar clusters for subject H04 is shown in Fig.~\ref{fig:alveolar_cluster_model}a, where each alveolar cluster element is visualized as a sphere with a volume equivalent to the gas volume as extracted from the CT image.
This alveolar cluster model is based on a four-element Generalized-Maxwell model,
which has been developed in \cite{ismailCoupledReducedDimensional2013} to describe the effective, visco-elastic behavior of alveolar tissue.
Specifically, it relates the flow rate of air into an alveolar duct (see Fig.~\ref{fig:alveolar_cluster_model}b) to the prevailing pressure difference between acinar pressure ($P_{\mathrm{a}}$) and pleural or inter-acinar pressure ($P_{\mathrm{pl/intr}}$) through several Maxwell elements consisting of springs and dashpots (see Fig.~\ref{fig:alveolar_cluster_model}c).
Each alveolar cluster is assumed to be comprised of a group of alveolar ducts as shown in Fig.~\ref{fig:alveolar_cluster_model}b.
In previous studies, the main stiffness component~$E_1$ has been assigned linear \cite{ismailCoupledReducedDimensional2013}, double-logarithmic \cite{ismailCoupledReducedDimensional2013}, and Ogden-like \cite{rothComprehensiveComputationalHuman2017,bel-brunonNumericalIdentificationMethod2014,rauschMaterialModelLung2011,birzleConstituentspecificMaterialBehavior2019} constitutive behavior.
Each has been shown to accurately describe the overall lungs constitutive behavior for their different use cases, but for healthy patients with moderate tidal volumes during inhalation the simple approach suffices.
From the four-element Generalized-Maxwell model (Fig.~\ref{fig:alveolar_cluster_model}c), we only consider the linear dashpot with viscosity parameter~$B_\mathrm{a}$ and a linear spring with stiffness~$E_1$, which results in a linear Kelvin-Voigt model and has been shown to be sufficient for the small tidal volumes considered here~\cite{geitnerApproachStudyRecruitment2022} and reduces the computational cost.
Finally, it is important to note that our patient-specific model indeed includes one alveolar cluster unit at every distal end of the full, 16 generation tree of conducting airways, and thus render it a true whole lung model as opposed to many alternative approaches using truncated lung morphologies.

\subsubsection{Pleurae, chest wall and diaphragm}
Elastic recoil of the chest wall and diaphragm as well as gravitational forces are accounted for by means of a suitable external pressure boundary condition acting on the alveolar clusters.
This pressure boundary condition therefore depends on the current volume of the lung model as it induces the elastic recoil of the chest wall. The condition also encompasses a hydrostatic pressure component that depends on the weight of the lungs as determined from the CT image by means of a density and volume analysis.
Further algorithmic details regarding all aspects of the simulation methodology as well as some recent extensions can be found in \cite{ismailCoupledReducedDimensional2013,rothComputationalModellingRespiratory2017,geitnerApproachStudyRecruitment2022}.

\subsection{Particle transport and deposition model}
\label{sec:SI_particletransport}
We developed a novel one-way coupled approach to compute particle transport and deposition throughout the entire lungs that hinges on the combination of two novel algorithmic components.
First, the computation of particle transport and deposition in reduced-dimensional airways in a Lagrangian fashion through reconstruction of the forces on the particle from the transient reduced-dimensional flow field, and second, a novel approach to combine many pre-computed local-scale 3D CFPD models into a surrogate model that enables the efficient simulation of particle transport across (reduced-dimensional) airway bifurcations.
This unique approach enables the tracking of all particles from their seeding in the trachea to the respiratory zone during inhalation as well as exhalation and, for the first time, allows the prediction of aerosol deposition throughout the lungs at sub-millimeter scale.
To account for the particle deposition in the mouth-throat region, we have implemented an analytical filter model according to~\cite{stahlhofenIntercomparisonExperimentalRegional1989}.
While this simple analytical filter model has been shown to agree very well with state-of-the-art 3D CFPD simulations \cite{borojeniSilicoQuantificationIntersubject2023}, this part of our modeling approach can also be readily replaced with a patient-specific 3D CFPD model of the upper airways in future works.
Due to the focus on the novel whole lung model in this paper and the excellent validation results obtained already with this simple analytical filter model, we decided to leave this extension for future work.

\begin{figure}[htb]
	\centering
	\vspace{0.25cm}
	\begin{tikzpicture}[scale=0.6]
		\usetikzlibrary{calc}

		\draw [thick] (-6.5,1) -- (0,1) -- (2,2.7) -- (2.9,1.85) -- (1.261052,-0.2) -- (2.8,-2.8) -- (1.9,-3.4) -- (0,-1) -- (-6.5,-1) -- (-6.5,1);

		\draw (-5.5,1) arc(90:-90:1.6 and 1);
		\draw (-5.5,1) to (-5.5,-1);

		\draw [-stealth] (-5.5,0.0) to (-3.9,0.0);
		\draw [-stealth] (-5.5,0.33) to (-4.0,0.33);
		\draw [-stealth] (-5.5,0.66) to (-4.32,0.66);
		\draw [-stealth] (-5.5,-0.33) to (-4.0,-0.33);
		\draw [-stealth] (-5.5,-0.66) to (-4.32,-0.66);

		\draw (-3.4, 0.5) node (TextNode1) {${v}_\text{fluid}$};

		\draw [dashed, ultra thick, black!30] (0,-1) -- (0,1);
		\draw [dashed, ultra thick, black!30] (0,-1) -- (1.261052,-0.2);

		\draw [-latex, ultra thick] (-1.5,-0.5) -- (-0.2,-0.5);
		\draw[blue,fill=blue] (-1.5,-0.5) circle (.8ex);f

		\draw [-latex, ultra thick] (0.5,-0.7) -- (1,-1.4);
		\draw[black!30!green,fill=black!30!green] (0.48,-0.7) circle (.8ex);

		\draw[fill=black!20] ($(-7,-5) + (180:0.90)$) arc (180:225:0.90) -- ($(-7,-5) + (225:1.2)$) arc (225:180:1.2) -- cycle;
		
		\draw [thick] (-7,-5) circle (1.5);
		\draw [thick] (-7,-5) circle (1.2);
		\draw [thick] (-7,-5) circle (0.90);
		\draw [thick] (-7,-5) circle (0.60);
		\draw [thick] (-7,-5) circle (0.30);
		
		\draw [thick] (-7,-5) -- +(0:1.5);
		\draw [thick] (-7,-5) -- +(45:1.5);
		\draw [thick] (-7,-5) -- +(90:1.5);
		\draw [thick] (-7,-5) -- +(135:1.5);
		\draw [thick] (-7,-5) -- +(180:1.5);
		\draw [thick] (-7,-5) -- +(225:1.5);
		\draw [thick] (-7,-5) -- +(270:1.5);
		\draw [thick] (-7,-5) -- +(315:1.5);5

		\draw[blue,fill=blue] (-7.96,-5.4) circle (.7ex);

		\draw [thick] (-1.5,-3.8) ellipse (1 and 0.25);
		\draw [thick] (-0.5,-4.3) arc(0:-180:1 and 0.25);
		\draw [thick] (-0.5,-4.8) arc(0:-180:1 and 0.25);
		\draw [thick] (-0.5,-5.3) arc(0:-180:1 and 0.25);
		\draw [thick] (-0.5,-5.8) arc(0:-180:1 and 0.25);
		
		\draw [thick] (-0.5,-3.8) -- (-0.5,-5.8);
		\draw [thick] (-2.5,-3.8) -- (-2.5,-5.8);
		
		\fill[black!50] (-0.75,-4.2) circle (0.08);
		\fill[black!50] (-0.75,-4.7) circle (0.08);
		\fill[black!50] (-0.75,-5.2) circle (0.08);
		\fill[black!50] (-0.75,-5.7) circle (0.08);

		\draw [thick] (4,-5) circle (1.2);
		\draw[black!30!green,fill=black!30!green] (3.5,-5.2) circle (.8ex);

		\draw [-latex, semithick] (-0.2,-0.7) to [bend right=15] (-5.5,-3);
		\draw [-latex, semithick] (-5,-5) to [bend right=15] (-3,-5);
		\draw [-latex, semithick] (0,-5) to [bend right=15] (2.2,-5);
		\draw [-latex, semithick] (3.8,-3.2) to [bend right=15] (0.8,-0.8);

		 \draw [black!50](-9.3,2.5) -- (-9.3,-7);

		\draw [thick] (-12.5,-3.8) ellipse (1 and 0.25);
		\draw [thick] (-11.5,-4.3) arc(0:-180:1 and 0.25);
		\draw [thick] (-11.5,-4.8) arc(0:-180:1 and 0.25);
		\draw [thick] (-11.5,-5.3) arc(0:-180:1 and 0.25);
		\draw [thick] (-11.5,-5.8) arc(0:-180:1 and 0.25);
		
		\draw [thick] (-11.5,-3.8) -- (-11.5,-5.8);
		\draw [thick] (-13.5,-3.8) -- (-13.5,-5.8);
		
		\fill[black!50] (-11.75,-4.2) circle (0.08);
		\fill[black!50] (-11.75,-4.7) circle (0.08);
		\fill[black!50] (-11.75,-5.2) circle (0.08);
		\fill[black!50] (-11.75,-5.7) circle (0.08);

		\node at (-13,-1) {\includegraphics[width=2.4cm]{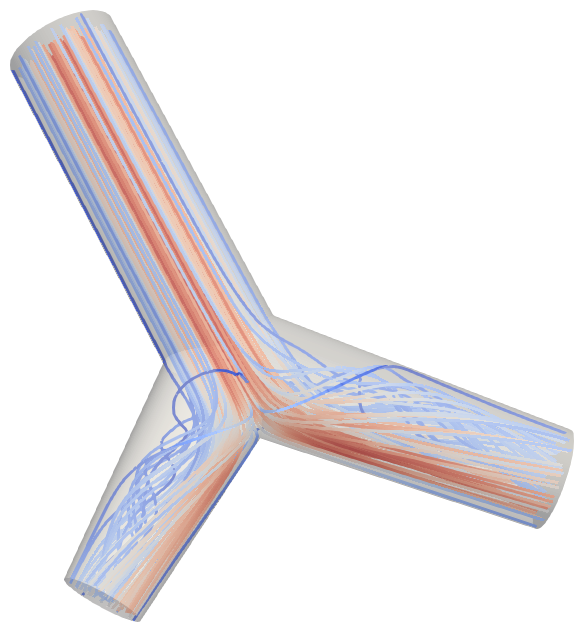}};
		\node at (-11.2,1) {\includegraphics[width=1.8cm]{fig_SI_particle_transfer_3d_flow_addon_01.png}};

		\draw [-latex, semithick] (-13,-2.3) to [bend right = 10] (-13,-3.5);
		\draw [-latex, semithick] (-11,0) to [bend left=40] (-12,-3.5);
	
	\end{tikzpicture}
	\caption{\textbf{Particle transport across airway bifurcations.}
	         Left: Schematic showing the creation of the particle transfer database based on a large number of three-dimensional computational fluid and particle dynamics (3D CFPD) simulations covering a large variety of airway geometries, flow regimes, and particle properties.
	         Right: Schematic showing a single reduced-dimensional airway bifurcation where first the particle exit position within the last cross-section of the parent airway (blue) is determined, second the pre-computed result of this particle transfer (green) is retrieved from the database, and finally the particle is transferred to this location within the first cross-section of the daughter airway.}
	\label{fig:databased_transfer}
	\vspace{0.25cm}
	
\end{figure}

The particles are modeled as point masses with spherical shape.
To simulate particle transport, we consider gravitational forces, flow resistance forces according to Reynolds, and a buoyancy force due to density differences between the particle and the fluid as external forces in Newton's second law of motion.
The resulting system of ordinary differential equations are solved using an explicit Forward-Euler time integration scheme.
To compute the forces on the particles resulting from the fluid flow, we leverage the instationary, reduced-dimensional flow field obtained from the patient-specific flow simulations to reconstruct the 3D fluid velocity field within each airway element.
This results in a velocity field which is transient in the amplitude, but steady in the profile across and along a single airway element for each time step.
An example for the parabolic flow velocity field at one cross-section ${v}_\text{fluid}$ is shown in Fig.~\ref{fig:databased_transfer} (right side).

Particle transport across airway bifurcations is computed using an interpolating surrogate model that is based on pre-computed local-scale 3D CFPD simulations of a representative airway bifurcation library.
Briefly, we conduct 3D flow simulations for a large library of airways bifurcations accounting for various flow regimes and geometries and subsequently simulate particle transport and deposition within these flow fields, again accounting for variations in a number of parameters such as particle density, size, as well as seeding location.
The behavior of particles flowing across these airway bifurcations is recorded, analyzed, and condensed into an interpolating surrogate model that is used to compute particle transport across airway bifurcations in the reduced-dimensional model as illustrated in Fig.~\ref{fig:databased_transfer}.
In the context of this work, the airway bifurcation library has been filled with a total of 315 scenarios, both systematically sampled from a large CT image database and including also the bifurcations from the first three airway generations of the in vivo reference data set from~\cite{conwayControlledParametricIndividualized2012}.

Particle deposition in the conducting airways is assumed to occur on contact of the particle with the airway wall.
In case the particle reaches the respiratory zone and enters an alveolar cluster with a given shape and volume, a deposition location within this alveolar cluster is chosen randomly.
Since our model encompasses the complete airway tree and lung tissue, particles never leave the simulation domain and hence our approach, in contrast to most other simulation approaches, can account for exhalation as well.
Therefore, particles, which do not get deposited inside the lung, will either be exhaled or stay in motion and can be simulated and tracked over consecutive breath cycles.
Moreover, this completeness of the lung model domain allows to state a total mass balance of the inhaled aerosol as exemplified in Fig.~\ref{fig:mass_balance_and_dep_per_airway_gen} and in turn enables detailed investigations such as comparing exhalation losses for different breathing patterns to name just one example.

It is important to emphasize that while the computation of the aforementioned bifurcation library is computationally expensive, this computation has only to be done once before it can be used in a variety of scenarios involving, e.g., different particle sizes, lung geometries and pathologies, and inhalation patterns.
Moreover, in combination with the reduced-dimensional transport and deposition simulation in the airway, this approach not only offers unprecedented levels of spatiotemporal tracking of particles in the whole airway tree and lung tissue, but is also efficient enough that patient-specific simulations of aerosol transport and deposition is now possible at scale.

\subsection{Simulation setup and boundary conditions}
\label{sec:SI_sim_setup_and_BCs}
The boundary conditions for the flow and particle simulations are straight forward, as they adhere as close as possible to the experimental setup described in~\cite{conwayControlledParametricIndividualized2012,katzControlledParametricIndividualized2013,majoralControlledParametricIndividualized2014,flemingControlledParametricIndividualized2015} and summarized in SI Sec.~\ref{sec:SI_exp_data}.
In the following the concrete conditions and simplifications are presented for the flow simulation and the one-way coupled particle simulation in turn.

\subsubsection{Flow simulation boundary conditions}
The respiratory system, as well as our simulation model of the lung, is composed of upper airways, lower airways, and respiratory zone, which are partially enclosed in a chest cavity.
The upper airways are omitted from the flow simulations, as the transport of particles through the upper airways is realized through an analytical filter model \cite{stahlhofenIntercomparisonExperimentalRegional1989} instead (see SI Sec.~\ref{sec:SI_model_methods}\ref{sec:SI_lungmodelflow} and \ref{sec:SI_model_methods}\ref{sec:SI_particletransport} for details).
All flow inlet boundary conditions are hence applied directly to the beginning of the trachea as can be seen in the supplementary videos.

In principle the lung model supports the simulation of spontaneous breathing by applying environmental pressure to the trachea and prescribing the chest cavities expansion and contraction through the volume directly or indirectly through the force of the muscles during breathing.
However, for the experiments we are reproducing, an AKITA device has been employed for enforcing a specific flow rate over time \cite{majoralControlledParametricIndividualized2014}.
This makes the simulation setup analog to that of mechanically ventilated patients for which the flow rate at the trachea is applied also.
No additional boundary conditions are needed in this case, as the inflation of the lung/alveolar clusters and the chest cavity, coupled through the pleural pressure, exert counteracting forces (see SI Sec.~\ref{sec:SI_model_methods}\ref{sec:SI_lungmodelflow} for details).
The magnitude of these forces depends on the material properties of the alveolar cluster elements and chest wall and is described in SI Sec.~\ref{sec:SI_model_methods}\ref{sec:SI_material_param_determination}.

\begin{figure}[htb]
  \begin{center}
    \textbf{a}
    \begin{minipage}[t][3.8cm][c]{0.35\textwidth}%
      \centering
      \includegraphics[width=0.9\textwidth]{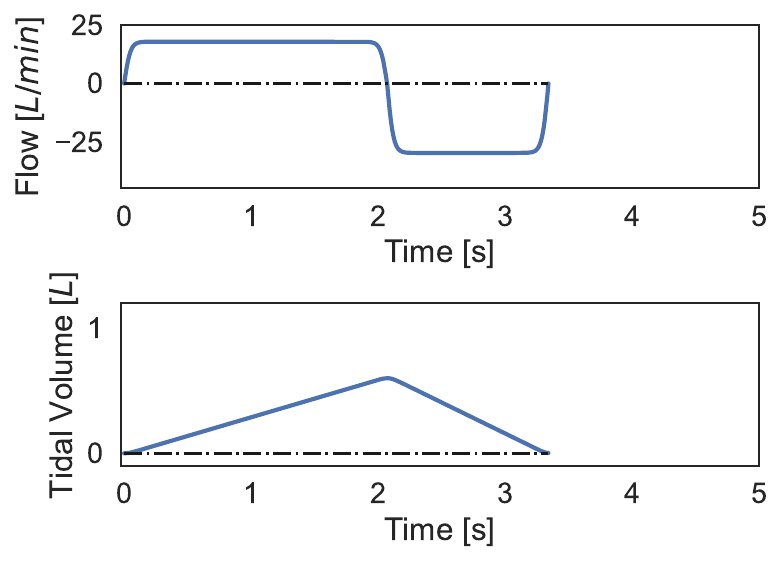}
    \end{minipage}%
    \hspace{1em}
    \textbf{b}
    \begin{minipage}[t][3.8cm][c]{0.35\textwidth}%
      \centering
      \includegraphics[width=0.9\textwidth]{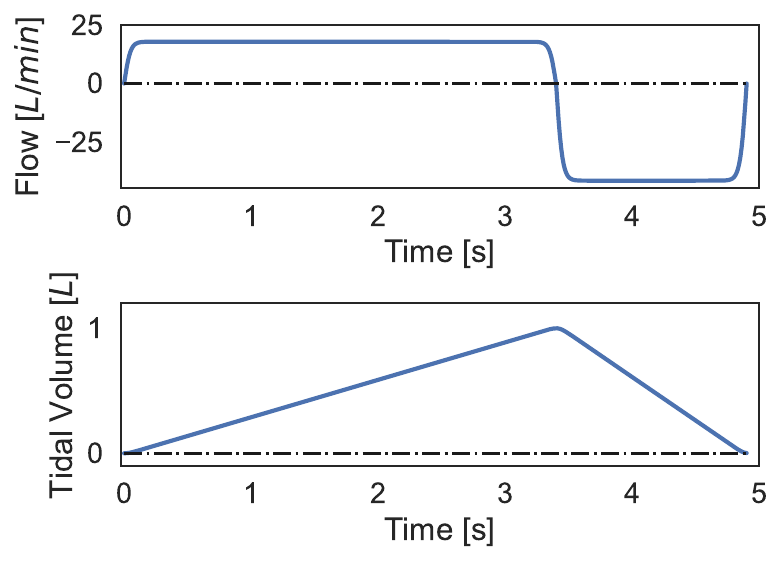}
    \end{minipage}%
  \end{center}
  \caption{\textbf{Breathing patterns.}
           Two different breathing patterns applied as flow rate boundary conditions in the simulations, closely following the setup of the experimental study (see~\cite{conwayControlledParametricIndividualized2012} and the summary in SI Sec.~\ref{sec:SI_exp_data}).
           \textbf{a} Shallow breathing with \qty{600}{\milli\liter} tidal volume.
           \textbf{b} Deep breathing with \qty{1000}{\milli\liter} tidal volume.
           }
  \label{fig:flow_BCs}
\end{figure}

The prescribed inhalation and exhalation flow rates are constructed from two regularized constant flow rates, as shown in Fig.~\ref{fig:flow_BCs}.
The inhalation part of the profile represents an idealized AKITA device, that applies a constant flow rate of \qty{18}{\liter/\minute}~\cite{majoralControlledParametricIndividualized2014} to the trachea.
Depending on which of the two cases is simulated (see Table~\ref{tbl:cases}), the inflow is applied for \qty{2.08}{\second} up to \qty{600}{\milli\liter} tidal volume or for \qty{3.41}{\second} up to \qty{1000}{\milli\liter} tidal volume.
In either case the first and last \qty{0.4}{\second} are regularized with a sigmoid to approximate the natural, smooth increase/decrease from/to zero flow and avoid unphysiological pressure peaks within the lung.
In the experiments the forced inhalation is followed up by a breath hold and exhalation that are done within certain time bounds, but at the patients discretion.
Diagrams in \cite{conwayControlledParametricIndividualized2012} indicate approximately constant exhalation rates and also some exhalation during the short breath hold phase that was needed to allow the study participants to switch from inhalation device to exhalation device, but is not considered to have a noticeable influence on deposition results.
Hence the breath hold and exhalation are subsumed under a single constant outflow rate with regularization for the simulated exhalation phase.
In \cite{conwayControlledParametricIndividualized2012} timings are provided for the exhalation phases at the two tidal volumes.
Averaging within each group leads to a simulated exhalation duration of \qty{1.27}{\second} at \qty{\sim 500}{\milli\liter/\second} and \qty{1.50}{\second} at \qty{\sim 700}{\milli\liter/\second} for \qty{600}{\milli\liter} and \qty{1000}{\milli\liter} tidal volume respectively.
A pressure increase and later decrease at the trachea inlet arises automatically from the lung tissue and chest cavity counteracting the expansion.
However, as the particle transport is driven mostly by the flow velocity, the pressure gradients along the airways are considered to  not impact the particle deposition, at least at these low pressure levels.

\subsubsection{Particle transport simulation boundary conditions}
\begin{figure}[htb]
  \begin{center}
    \textbf{a}
    \begin{minipage}[t][4.5cm][c]{0.4\textwidth}%
      \centering
      \includegraphics[width=0.9\textwidth]{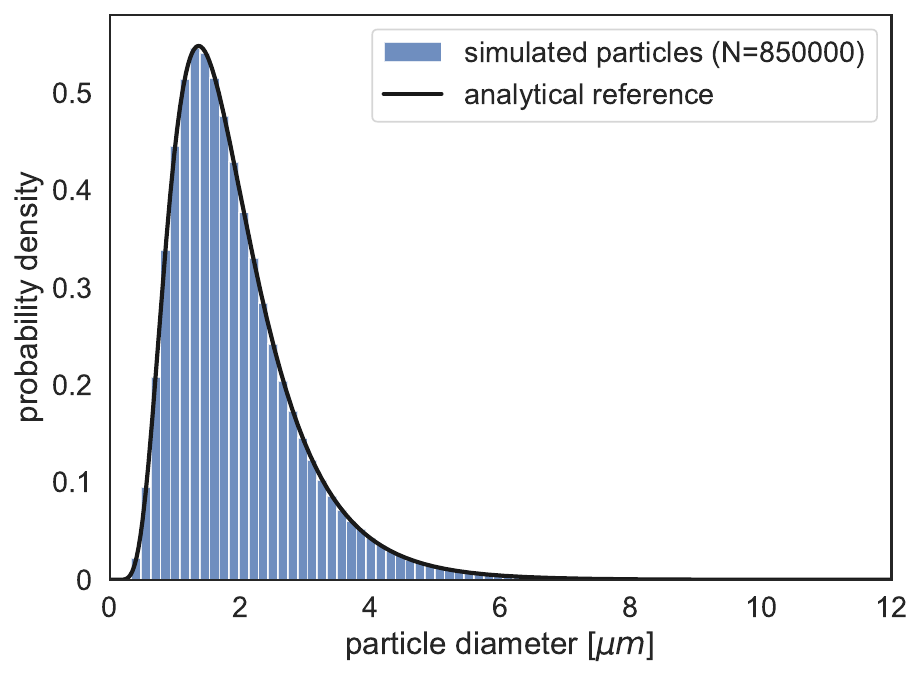}
    \end{minipage}%
    \hspace{1em}
    \textbf{b}
    \begin{minipage}[t][4.5cm][c]{0.4\textwidth}%
      \centering
      \includegraphics[width=0.9\textwidth]{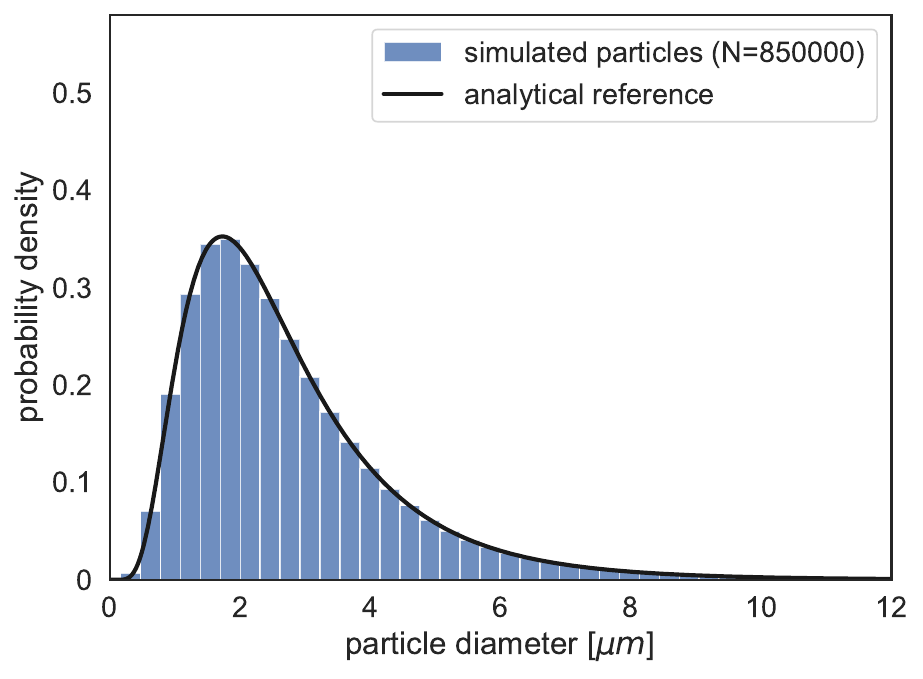}
    \end{minipage}%
  \end{center}
  \caption{\textbf{Aerosol particle sizes.}
           Two different aerosol particle size distributions used for the particle simulations (blue), reproducing exactly the setup and measurements of the experimental study cases (see~\cite{conwayControlledParametricIndividualized2012} and the summary in SI Sec.~\ref{sec:SI_exp_data}).
           For comparison, the analytical reference defined as a log-normal distribution with matching mass median aerodynamic diameter (MMAD) and geometric standard deviation (GSD) is plotted as black solid line.
           \textbf{a} Small aerosol size with MMAD of \qty{3.38}{\micro\meter} and GSD of \qty{1.61} as specified for case 2B.
           \textbf{b} Large aerosol size with MMAD of \qty{6.05}{\micro\meter} and GSD of \qty{1.75} as specified for case 3B.
           }
  \label{fig:particle_size_distributions}
\end{figure}

As for the flow simulation, the boundary conditions to the particle transport simulation strive to reproduce the experiments from~\cite{conwayControlledParametricIndividualized2012} and summarized in SI Sec.~\ref{sec:SI_exp_data} as close as possible.
Hence the particles are seeded statistically over time into the flow during inhalation.
The distribution of particle sizes seeded in the simulation matches the MMAD and GSD as provided in Table~1 of \cite{katzControlledParametricIndividualized2013}.
Before injection into the trachea however, the particles are filtered through an analytical filter function~\cite{stahlhofenIntercomparisonExperimentalRegional1989} to account for the deposition in the upper airways (see SI Sec.~\ref{sec:SI_model_methods}\ref{sec:SI_particletransport} for details).
The resulting particle size distributions for a large and a small aerosol size are shown in Fig.~\ref{fig:particle_size_distributions}.
Comparing the randomly generated sample of particles to the analytical reference of a log-normal distribution with matching MMAD and GSD shows that both are in excellent agreement for the used number of particles~$N$.
The particles are seeded at a constant rate of 250,000 particles per second during the inhalation phase, thus leading to a total number~$N$ of 520,000 and 850,000 particles for the cases with shallow and deep breathing pattern, respectively.
Refer to Table~\ref{tbl:cases} for an overview of the setup used for each of the ten cases.
The particles are released into the flow uniform over the area of a plane at the beginning of and perpendicular to the trachea.
This emulates the mixing of the particle jet from the glottis by the turbulence within the trachea, but which is not resolved by the reduced-dimensional flow.

\subsection{(Material) parameter determination and calibration}
\label{sec:SI_material_param_determination}
The patient specific lung models must be armed with a broad set of material and algorithmic parameters.
The source of these parameters range from calibration using patient individual measurements to population average from literature, mainly depending on the availability of data and the underlying modeling assumptions.

\subsubsection{Reduced-dimensional whole lung model}
Determining patient specific material parameters for the reduced-dimensional lung model is challenging due to the large number of components that make up the human respiratory system and their tight interactions.
Naturally, the model should be limited in it's complexity to the problem it is intended to solve, which is the accurate description of flow distribution in this case.
The particle deposition is considered to be driven primarily by the flow velocity distribution, which in turn depends primarily on the individual lung and airway tree geometry and much less on the constitutive behavior.
Due to the healthy and homogenous state of the lungs in the conducted validation study, the distribution of stress, strain, and therefore also flow to the alveolar clusters is uniform within the tidal volumes prescribed by the experiments.
We expect that the actual levels of stress and strain become important for checking the internal consistency and the application to extreme inhalation maneuvers or patients with lung pathologies.
The latter is briefly outlined in the section on \textit{Tissue disease modeling} of the main text, but not relevant for the healthy subjects of the validation study.
For the following fitting procedure of the material parameters to the experimental measurements a steady single compartment version (growth stopped after first generation) of the lung model introduced in SI Sec.~\ref{sec:SI_model_methods}\ref{sec:SI_model_gen} is assumed, from which the properties are then distributed to the whole lung model.
The system of equations for the single compartment is solved under the following series of conditions.
The first data point is the CT into which the model is grown and that corresponds to a mean tidal breath-hold \cite{conwayControlledParametricIndividualized2012}.
The gas volume within the lungs at that state is recovered from the Hounsfield units (\unit{\hounsfield}) assigned to the CT voxel within the lung mask, which are taken as a binary mixture of tissue (\qty{0}{\hounsfield}) and gas (\qty{-1000}{\hounsfield}) \cite{gattinoniWhatHasComputed2001}.
The next data point is the functional residual capacity (FRC) that was measured in a standing position for every participant.
As the FRC is approximately \qty{40}{\percent} larger in a standing rather than a supine position \cite{ibanezNormalValuesFunctional1982}, the FRC in the supine position of the CT can be approximated.
In the relaxed state at FRC there is an equilibrium between contractile forces of lung tissue and tensile forces of the chest cavity which results in a negative pressure in the pleural cavity \cite{gattinoniTargetingTranspulmonaryPressure2019}.
For this mean pleural pressure typically a range of $-3$ to \qty{-5}{\centimeterwater} is stated in literature, with one concrete measurement at \qty{-4.11}{\centimeterwater} \cite{zielinska-krawczykPleuralManometryHistorical2018,aronIntrapleuraleDruckBeim1900}.
The mean pleural pressure is further supplemented by a spatially varying component that accounts for gravity through a linear hydrostatic pressure profile over the height dimension, determined from the individual's average lung density and thus weight.
The specific elastance of the lung is taken as \qty{13.8}{\centimeterwater}, which is an average of the non-ARDS cases published in \cite{chiumelloLungStressStrain2008}.
The damping element of the Kelvin-Voigt model is assigned a viscosity of \qty{32.9}{\centimeterwater \second/\liter} following \cite{rothCouplingEITComputational2017,ismailCoupledReducedDimensional2013}. 
Another contribution to the overall stiffness of the respiratory system is the chest cavity, comprised of chest wall and diaphragm with abdomen.
They are subsumed under a single chest wall elastance, which is determined through the lung elastance to total respiratory elastance ratio.
This ratio was approximated at $0.71$ for non-ARDS patients in the measurements provided by \cite{chiumelloLungStressStrain2008}.
The specific chest wall elastance is then \qty{5.7}{\centimeterwater}.
With the conditions specified above, the single compartment surrogate of the whole lung model can be solved for the volume of the lungs and chest cavity in a (fictitious) stress free state.
They are essential in specifying the patient specific stress strain relationship.
The lung model parameters determined so far apply to a patient in supine position.
However, the inhalation experiments were carried out in an erect position.
Under the simplifying assumption that the change in position only affects the chest cavity's behavior, the FRC measured for standing study participants can be used to determine a new system state \cite{conwayControlledParametricIndividualized2012}.
As the difference between supine and erect positions is a change in the direction of the constant gravitational pull, the change in the chest wall condition is taken to be a constant force term as well.
Solving for this constant offset in the pleural pressure, a patient specific, standing adjusted specific elastance and stress-free reference volume of the chest cavity is obtained.

\subsubsection{Gas and particle characteristics}
A further required set of parameters for solving the flow problem in the human lungs are the gas and particle properties.
Particles are considered to be spherical droplets of an aquaeous solution.
The flow medium is air in an incompressible regime at a density of \qty{1.18}{kg/m^3} and dynamic viscosity of \qty{17.9e-8}{\centimeterwater/\second} in accordance with \cite{rothCouplingEITComputational2017}.
The effects of a temperature or humidity increase are not accounted for.

\section{Additional information on post-processing and data analysis}
\label{sec:SI_post_processing}
The main output of the simulation model described in the previous sections are the spatially resolved flow and pressure distribution throughout the lungs as well as the precise location and velocity of each aerosol particle, all in a temporally highly resolved manner.
While the former two can be used to derive space and time-dependent ventilation and strain maps, the latter can be used to gauge and quantify aerosol deposition throughout the entire lungs at an unprecedented spatial resolution.
The ability to track every single particle trajectory over time even allows to study the dynamics of aerosol transport and deposition over the course of a breath cycle for example, which goes far beyond the usual assessment of the final deposition pattern after finishing the inhalation.
Although it is not used in the current validation study due to the lack of in vivo reference data, we would like to emphasize this new opportunity and showcase the high temporal resolution of the simulation result data in the videos provided as supplementary electronic files in Videos~\ref{vid:1A}--\ref{vid:6A}.

\subsection{Deposition fraction per lung and per lobe}
On the simulation side, the precise deposition location of every particle is known, so using the segmentation masks for the left/right lung and each lobe, it is straightforward to compute the fraction of aerosol mass deposited in each of these sub-domains.
On the experimental side, however, we have to deal with the fact that the high-resolution CT (HRCT) image and the SPECT/CT image are not aligned, and in turn also the segmentation masks obtained from the CT image are not directly applicable to the SPECT/CT image data.
Hence, to be able to compute deposition fractions from the SPECT/CT images, we first aligned the HRCT image and SPECT/CT image using deformable image registration techniques.
Using these deposition fractions from the SPECT/CT images as a reference, we were able to validate the predicted deposition fractions from the simulation by means of direct comparison, correlation analysis or suitable error metrics (see SI Sec.~\ref{sec:SI_post_processing}\ref{sec:SI_error_metrics} for details).

\subsection{Central versus peripheral deposition}
\label{sec:SI_c2p_split_analysis}
Analyzing the deposition fraction in peripheral versus central parts of the lungs is commonly used to gauge the fraction of particles that end up in smaller, high-generation airways and the respiratory zone as compared to those depositing in the larger airways.
We followed the approach described in \cite{perringNewMethodQuantification1994,flemingValidationConceptualAnatomical2004,flemingComparisonSPECTAerosol2006} to divide the lungs into ten concentric shells and to consider the five inner shells as the central region and the five outer shells as the peripheral region.
This is done individually for the left and the right lung, such that we end up with four spatial regions (left-central, left-peripheral, right-central, right-peripheral) that can again be used to validate the predicted deposition fractions against the in vivo reference data from the SPECT/CT image as described for lobar deposition fractions in the previous section.
See Fig.~\ref{fig:cp_deposition}a for a 3D visualization of the four regions obtained for subject H04.

\subsection{Deposition in conducting airways versus respiratory zone}
The simulation results allow the direct assessment of the fraction of aerosol mass deposited in the conducting airways as compared to the respiratory zone.
This is not possible for the in vivo reference data due to the limited spatial resolution of SPECT/CT and even HRCT images.
Therefore, we used the 24h clearance obtained from scintigraphy as a proxy for the conducting airway deposition fraction (CADF) and analyze the correlation with the predicted CADF from the simulation in Fig.~\ref{fig:cadf}.
To showcase the even higher spatial resolution and result data quality on the simulation side, we also compute the deposition fraction for each generation of the airway tree as shown in Fig.~\ref{fig:mass_balance_and_dep_per_airway_gen}.

\subsection{Volume normalized deposition fraction}
Considering the comparison of deposition in healthy versus diseased parts of the lung, it is necessary to use a volume-normalized deposition fraction, because directly comparing the deposition fractions is inconclusive due to the fact that the two sub-domains (healthy/diseased) differ in size.
Therefore, we divide the deposition fraction by the volume fraction of the respective sub-domain in order to get a volume normalized measure that allows to judge the effect of disease on the deposition pattern.
Note that this measure is equivalent to a normalized deposition density we obtain from dividing the average deposition density (deposited aerosol mass/volume) in the sub-domain by the average deposition density in the entire domain, i.e., lungs.

\subsection{Correlation and error metrics}
\label{sec:SI_error_metrics}
To quantify the correlation between predicted deposition fractions from simulation and the corresponding reference values from the experimental study, we compute Pearson's correlation coefficient~$\rho$ for all available data points, i.e., every sub-domain and all 10 inhalation experiments.
See Figs.~\ref{fig:lobar_deposition}d, \ref{fig:cp_deposition}d, \ref{fig:cadf}b, and \ref{fig:per_lung_deposition}d for the respective number of data points used in the correlation analysis and corresponding scatter plot.
To quantify the quality of an individual model prediction, we defined the error as the mass fraction of deposited aerosol that is misattributed to the considered sub-domain, e.g., the lower left lobe for one specific case 4A.
On the ensemble level, we use a box-and-whisker plot showing the median (orange line), the range between 25th and 75th percentile (box) and the minimum/maximum value (whiskers) overlaid by a scatter plot showing all individual data points (blue) with some randomly added jitter along the horizontal axis for visibility.
See Figs.~\ref{fig:lobar_deposition}d, \ref{fig:cp_deposition}d, and \ref{fig:per_lung_deposition}d for examples.

\clearpage

\section{Additional result data}
\label{sec:SI_result_data}

\subsection{Validation of predicted deposition fraction per lung}
\begin{figure}[htb]%
  \centering
  \includegraphics[width=0.9\textwidth]{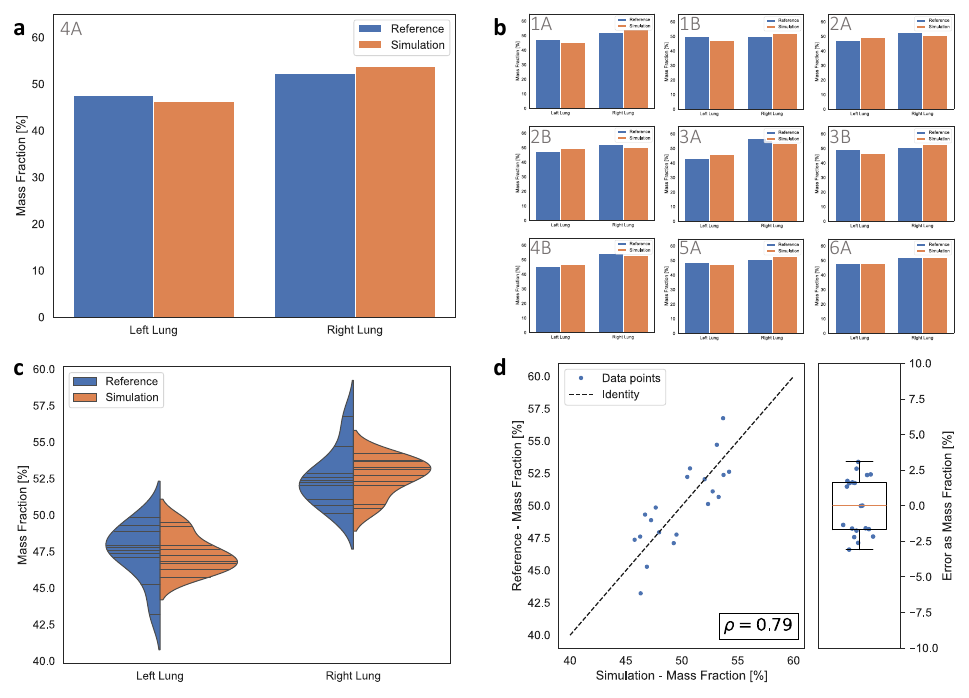}
  \caption{\textbf{Validation of predicted aerosol deposition per lung.}
      \textbf{a} Deposited aerosol mass as a fraction of total aerosol mass deposited in the lungs. Comparison of model predictions (orange) and in vivo reference data (blue) for subject H04 (visit A).
      \textbf{b} Comparison of deposition fractions per lung for the remaining 9 inhalation experiments.
      \textbf{c} Distribution of model predictions (orange) and reference values (blue) for all 10 inhalation experiments (black lines indicate individual cases).
      \textbf{d} Scatter plot showing correlation and box plot showing error values for all 20 data points (10 inhalation experiments and left/right lung), respectively.
          The error is defined as the mass fraction of deposited aerosol that is misattributed.}
  \label{fig:per_lung_deposition}
\end{figure}

Fig.~\ref{fig:per_lung_deposition} summarizes the validation results for the predicted aerosol deposition fraction per lung.
The analysis and interpretation is analogous to the one for lobar deposition fractions (see Fig.~\ref{fig:lobar_deposition} in the main text).

\movie{Particle simulation of one full breathing cycle for case 1A.}\label{vid:1A}
\movie{Particle simulation of one full breathing cycle for case 1B.}
\movie{Particle simulation of one full breathing cycle for case 2A.}
\movie{Particle simulation of one full breathing cycle for case 2B.}
\movie{Particle simulation of one full breathing cycle for case 3A.}
\movie{Particle simulation of one full breathing cycle for case 3B.}
\movie{Particle simulation of one full breathing cycle for case 4A.}
\movie{Particle simulation of one full breathing cycle for case 4B.}
\movie{Particle simulation of one full breathing cycle for case 5A.}
\movie{Particle simulation of one full breathing cycle for case 6A.}\label{vid:6A}

\bibliography{ebenbuild}